**Atomically Thin Quantum Spin Hall Insulators**

*Michael S. Lodge, Shengyuan A. Yang, Shantanu Mukherjee, and Bent Weber\**

Dr. M. S. Lodge
School of Physical and Mathematical Sciences, Nanyang Technological University, 637371, Singapore

Prof. S. A. Yang
Research Laboratory for Quantum Materials, Singapore University of Technology and Design, 487372 Singapore

Prof. S. Mukherjee
Department of Physics, Indian Institute of Technology Madras, Chennai, 600036, India

Prof. B. Weber
School of Physical and Mathematical Sciences, Nanyang Technological University, 637371, Singapore
Australian Research Council (ARC) Centre of Excellence for Low-Energy Electronics Technologies (FLEET), School of Physics, Monash University, Clayton VIC 3800 Australia
E-mail: b.weber@ntu.edu.sg

Atomically thin topological materials are attracting growing attention for their potential to radically transform classical and quantum electronic device concepts. Amongst them is the quantum spin Hall (QSH) insulator – a two-dimensional state of matter that arises from an interplay of topological band inversion and strong spin-orbit coupling, with large tunable bulk band gaps up to 800meV and gapless, one-dimensional edge states. Reviewing recent advances in materials science and engineering alongside theoretical description, this article surveys the QSH materials library with focus on their prospects for QSH-based device applications. In particular, this review discusses theoretical predictions of non-trivial superconducting pairing in the QSH state towards Majorana based topological quantum computing – the next frontier in QSH materials research.

## 1. Introduction

Since its first isolation in 2005, graphene – a single atomic layer of carbon atoms arranged in a honeycomb lattice (as illustrated in **Figure 1**) is often regarded the prototypical two-dimensional (2D) material.[1, 2] It has become synonymous with atomic-thinness, high tensile strength [3, 4], and electrical mobility.[5, 6] More importantly, it has also become a fertile hunting ground for unconventional electronic behavior, such as manifestations of "massless" Dirac fermions,[1] and unconventional Hall responses.[1, 7-14] The graphene lattice was also the first material in which a time-reversal symmetry (TRS) protected topological insulating state in two dimensions[15, 16] was predicted – the quantum spin Hall (QSH) insulator.[15, 16]

Different from trivial semiconductors or insulators, in which valence and conduction bands are strictly separated by a (trivial) electronic band gap, 2D topological insulators (2D TIs) have a gapped 2D bulk, but possess one or more gapless electronic states confined to their boundary. In the QSH insulator (Figure 1A,B), a subclass of 2D TIs, the boundary harbors a pair of symmetry protected, spin-momentum locked one-dimensional (1D) boundary modes. This can be understood as the result of the orbital structure of certain crystal lattices similar to that of graphene, giving rise to an inversion of conduction and valence electron bands. Such band inversion in combination with spin-orbit coupling (SOC) leads to a net spin accumulation transverse to an applied electric field, which results in a quantized spin Hall voltage at zero net charge Hall voltage; hence, the quantum spin Hall effect.

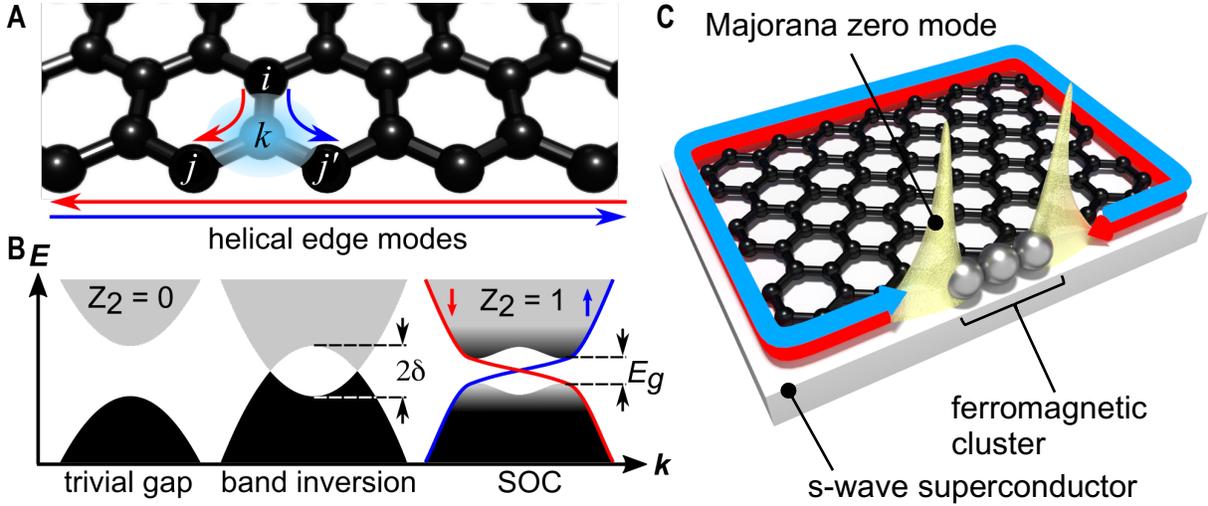

**Figure 1. Electronic properties of Quantum Spin Hall Insulators.** (A) Honeycomb lattice illustrating Kane-Mele spin-orbit coupling. Spin-dependent, next nearest neighbor hopping results in the formation helical (spin-momentum locked), one-dimensional conduction modes (red and blue curves) that counter-propagate at the QSH edges. (B) Schematic band diagram of conduction and valence bands (gray and black respectively), comparing a trivial insulator ($Z_2 = 0$) and a semimetal having linearly-dispersing (Dirac) nodes, resulting from band inversion with inversion gap $2\delta$. Spin-orbit coupling opens a band gap ($E_g$) within the 2D bulk and gives rise to linearly-dispersing, spin-momentum locked modes (red and blue) localized at the edges ($Z_2 = 1$). (C) Breaking time-reversal symmetry locally by, for example, magnetic or exchange fields could lead to the formation of Majorana bound states at the helical edge if superconductivity is induced by proximity to an *s*-wave superconductor.

Although the first experimental confirmation of the QSH state was realized in semiconductor heterostructures of HgTe/CdTe[17-19] and InAs/GaSb,[20-23] the iconic graphene lattice has remained a key model system for atomically thin QSH insulators. The recent emergence of atomically thin quantum spin Hall materials, with a plethora of related hexagonal lattice structures and strong spin-orbit coupling,[15, 16, 24-37] provides an ever growing, rich pool for scientific discovery and potential for technological applications in electronics, spintronics, and even quantum information processing.

Electrons within the 1D QSH edge modes (Figure 1B) are spin-momentum locked (*helical*) such that each spin polarity is tied to one momentum direction. Electronic states are thus protected against scattering from non-magnetic disorder by time-reversal symmetry.[15, 24, 26, 27, 34, 35, 38] QSH based devices may be robust against localization even by strong disorder,[34] edge roughness,[39, 40] or non-magnetic impurities.[26, 27] As spin and momentum are locked along the helical edge, QSH insulators may further be used for the generation and detection of spin currents by spin and charge interconversion[41, 42] in spintronic devices.[41] Finally, controlling the topological phase itself by electric[40, 43-45] and magnetic fields, or by strain,[43, 46] may offer control over charge and spin currents at the helical edge with rapid switching rate[47] compared to the charge-based accumulation and depletion underlying complementary metal oxide semiconductor field effect transistor (MOSFET) devices.

One key requirement to harness the rich physics of QSH materials towards prospective electronic device applications remains that the size of the topological band gap $E_g$ must be substantially larger than the thermal broadening at the operating temperature, $T$ ($E_g \gg 3.5\ k_B T$, where $k_B$ is the Boltzmann constant). For room temperature operation, this will require a gap magnitude significantly larger than 90 meV, which exceeds that estimated for graphene by 3 orders of magnitude.[48-50] In Xenes – materials with graphene-like hexagonal lattice structure - the magnitude of $E_g$ is primarily governed by the strength of the spin-orbit coupling (SOC), which motivates an ongoing search for materials with related hexagonal lattice structures but larger

SOC. In addition, interface effects, lattice strain and distortions, as well as electric fields can have a profound influence on the band structure, including the band inversion itself and the magnitude of the topological gap. A detailed discussion of the role of SOC in QSH materials is included in Section 3, while we discuss the effects of strain and electric field in the context topological phase transitions in Section 7. A substantial library of QSH materials with topological gaps ranging between $E_g \approx 40$ μeV[51] to 800 meV[52] has now been realized, whose materials' details and characterization we review in Section 5. The recent discovery of atomically flat bismuthene sets the record of the largest QSH gap with $E_g \approx 800$ meV[52], shortly followed by $Na_3Bi$ with an electric field tunable gap of 0-300 meV[53] promising applications approaching room temperature. The highest temperature in which the QSH state has been demonstrated in a device is 100 K in atomically thin monolayers of $WTe_2$.[54]

It seems a matter of time until a QSH state is demonstrated at room temperature, forming a viable route towards QSH-based electronic devices[40, 43-46, 54, 55] such as "topological field effect transistors" with fast switching rate and low-dissipation. We review proposals and recent experimental demonstrations of the control of the topological phase in Section 7.2. Meanwhile, at the extreme of low temperature, the interplay of topology and superconductivity[56-58] is predicted to lead to exotic quasiparticle excitations,[56, 59] including non-Abelian anyons. The much-coveted Majorana fermion, for example has known implications for fault-tolerant quantum information processing.[56, 57, 60] We review related developments in Section 7.4.

**2. History of the Quantum Spin Hall State**

The distinct properties of 2D TIs can be classified in the broader concept of topology – the study of properties which are preserved under smooth deformations – whereby topological properties are characterized by integer numbers, the so-called topological invariants.

Historically, characterizing the electronic properties of condensed matter systems by topology was first applied in 1982[61] to a two-dimensional electronic gas in a strong magnetic field in order to explain the Landau Level quantization in the integer *quantum Hall effect* (QHE)[62]. The corresponding topological invariant is the Chern number $n \in \mathbb{Z}$, where $\mathbb{Z}$ is an integer, quantifying the total Berry flux in the Brillouin zone. Via the *bulk-boundary correspondence*,[63] the Chern number quantifies the number of states localized at the boundary between bulk and vacuum.

These ideas were further developed by Haldane[64] who considered a 2D crystal lattice with honeycomb lattice structure similar to graphene, but in which each atom has an on-site magnetic moment of opposite polarity to its neighbors such that the net macroscopic magnetic field was zero while the system remained topologically nontrivial. In the Haldane model, electrons may "hop" around the perimeter of the honeycomb lattice in one preferred direction, giving rise to a non-zero charge Hall voltage that is quantized. The result is equivalent to the QHE, but may exist in the absence of an applied magnetic field. This effect is hence known as the *quantum anomalous Hall effect* (QAHE).

The first realization of a material in which the QSH state was predicted was graphene. Within one year of its experimental isolation in monolayer and bilayer forms by Novoselov et al.[1, 2] in 2004-2005, Kane and Mele[15] described the topology of this natural realization of a 2D honeycomb lattice in a manner similar to that of Haldane's model, but in which spin-orbit coupling takes the role of the alternating magnetic order. As such, two copies of the Haldane model were naturally realized giving rise to a vanishing charge Hall voltage but a non-zero spin Hall voltage. In this *quantum spin Hall* state, similar to the QHE, two counterpropagating states are localized at the edges. However, these states are spin-momentum locked such that different spin polarities have opposite direction of momentum, as illustrated in Figure 1B. Unlike the QHE or the QAHE, which break time-reversal symmetry due to the respective magnetic field or magnetic moments, QSH insulators inherently require time-reversal symmetry for its protection[16, 25]. Due to the difference in the defining symmetry, the QSH insulator is described not by Chern number, but in terms of a $Z_2$ topological invariant.[16] Odd ($n = 1$) or even ($n = 0$) value of the $Z_2$ invariant indicates, respectively, whether a material possesses nontrivial topology or not.

Despite the immense importance of the graphene lattice in the development of the topological classification of band structures, it was soon realized that the weak spin-orbit (SO) coupling inherent to graphene would lead to a SO gap so small that bulk conduction channels would thermally activate at any reasonable experimental temperatures, precluding observation of QSH signatures.[48-50, 65, 66] In the search for materials with stronger spin-orbit coupling, Bernevig, Hughes, and Zhang subsequently made predictions[24] for the QSH state in semiconductor heterostructures of HgTe/CdTe, that were soon confirmed experimentally by König et al. in 2007.[17] Later in 2011, a QSH state was also confirmed in the related heterostructure of InAs/GaSb by Knez et al.[22]

In these carefully designed semiconductor heterostructures, band inversion arises from an intricate balance of thickness control and doping of the constituent semiconducting layers,[17, 22, 24, 30, 67] with spin-orbit gap of up to 55 meV in compressively strained HgTe.[68] Many of the predicted properties of the QSH insulator have subsequently been experimentally observed in heterostructure-based QSH insulators, including the concurrence of bulk gap and edge states,[69] quantized 1D conduction at the edge,[17, 22] and the edge state's spin-polarization.[19] However, the comparatively small QSH gap (4 to 55 meV[22, 68]), and the fine balance of thickness control and doping required leaves semiconductor heterostructures susceptible to disorder,[70] which poses a challenge to both observation and application of the QSH state. This ultimately motivates an ongoing search of QSH materials having larger band gaps, and in which band inversion is an intrinsic property of their crystal lattice rather than by virtue of heterostructure design.

## 3. Size matters: The Role of The Spin-Orbit Coupling

The size of the QSH gap is critically determined by the strength of the spin-orbit coupling (SOC). SOC is the interaction between the spin and the orbital motion of an electron and is a purely relativistic effect. The coupling is inherent in the Dirac equation, and, when taking the non-relativistic limit to obtain the Pauli equation, SOC appears in the following explicit Hamiltonian form

$$H_{\text{SOC}} = \frac{\hbar}{4m^2c^2}\boldsymbol{\sigma} \cdot (\nabla U \times \boldsymbol{p}). \tag{1}$$

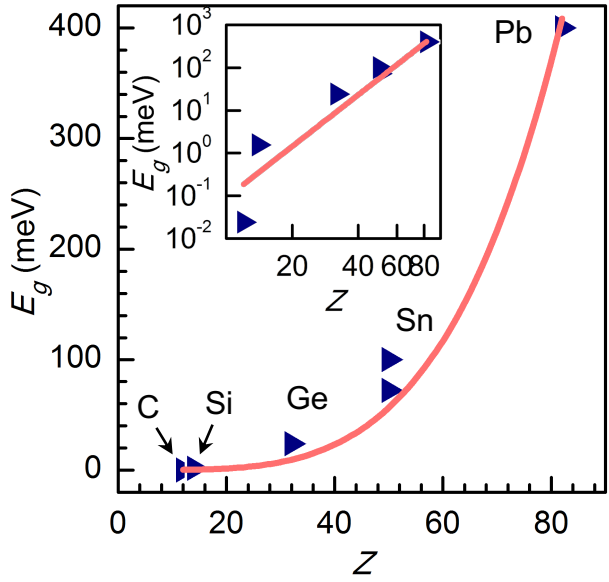

**Figure 2. Size of the topological band gap in group IVA Xenes.** (A) First principles calculation of the spin-orbit induced topological band gap in group IVA Xenes. The data is compiled from Ref's [49, 50, 106, 111]. The red line shows the expected the $Z^4$ dependence of atomic spin-orbit coupling in the Kane-Mele model. (Inset) $E_g$ vs $Z$ plotted on a double-logarithmic scale.

Here, $U$ is the potential energy of the electron. In atomic physics, the SOC is usually cast into the form of $H_{SOC} = \zeta \mathbf{L} \cdot \mathbf{s}$, where $\mathbf{L}$ and $\mathbf{s}$ are respectively the orbital and spin angular momenta, and the coupling strength $\zeta \propto \left(\frac{1}{r}\right)\left(\frac{\partial U}{\partial r}\right)$ with $U$ being the ionic potential. The SOC strength increases with atomic number $Z$. As a rough guide, the coupling strength scales as $\zeta \sim Z^4$, resulting from a stronger on-site Coulomb potential of the increased proton count in the atomic nucleus. In 5d elements, for example, $\zeta$ can reach about 0.5 eV.

In crystalline solids, SOC is still a small perturbation on the overall energy scale. However, since the physical properties are mostly determined by a small energy window around the Fermi level, when zooming into this window, SOC can play an important role in shaping the low-energy band structure. SOC thus becomes an important consideration for materials engineering. In **Figure 2**, we compare density functional theory (DFT) predictions of the QSH gap for hexagonal lattices of the group-IV elements, reflecting the approximate $\sim Z^4$ dependence.

In theoretical treatment, the physics are usually captured by low-energy effective models. In such models, SOC could take different forms depending on the symmetry of the system. For example, in 2D electron gas based on III-V semiconductor quantum wells, one often encounters the Dresselhaus SOC and the Rashba SOC terms, given by

$$H_D = \beta(k_x\sigma_x - k_y\sigma_y) \qquad (2)$$

and

$$H_R = \alpha(k_x\sigma_y - k_y\sigma_x), \qquad (3)$$

respectively. The Dresselhaus SOC is due to the broken inversion symmetry in the bulk lattice structure, whereas the Rashba SOC originates from the interface-induced asymmetry.

The Rashba type SOC also frequently appear in thin films and 2D materials from substrate effects. This can be readily observed from the original SOC Hamiltonian in Equation (1): the substrate generates an effective potential energy gradient normal to the 2D plane, thus $H_{SOC} \propto \boldsymbol{\sigma} \cdot (\hat{z} \times \boldsymbol{p})$, leading to the Rashba form of SOC. In the Kane-Mele model, one encounters the intrinsic SOC term. This term can be understood in the following picture. When hopping from a site $i$ to its second nearest neighbor $j$ on the honeycomb lattice, the electron experiences a potential gradient $\nabla U$ generated by the nearest neighbor $k$ site (see Figure 1A). This gradient is in-plane and normal to the hopping direction, thus the resulting SOC is $\propto \sigma_z$ and takes the form as in Equation (2).

|  | Theory | | Experiment | | | Refs. |
|---|---|---|---|---|---|---|
|  | Substrate | QSH gap [meV] | Substrate | QSH Gap [meV] | $E_F$ position [meV] |  |
| Semiconductor Heterostructures ||||||||
| HgTe/CdTe-Cd$_{0.5}$Zn$_{0.5}$Te |  |  | GaAs | ~55 |  | [68] |
| InAs/GaSb |  |  | AlSb | ~4 |  | [22] |
| Xenes ||||||||
| Graphene |  | ~0.001 – 0.050 |  |  | Mid gap | [48-50, 65, 66] |
| Silicene |  | 1.55 – 2.9 |  |  | Mid gap | [85] |
| Germanene |  | 23.9 |  |  | Mid gap | [85, 106] |
| Stanene |  | ~73, 100 |  |  | Mid gap | [90, 106, 112] |
| (decorated) |  | ~270 – 340 |  |  | Approx. mid gap | [90] |
| (ultra-flat) | Cu(111) | ~220 | Cu(111) | 300 | -1250 | [107] |
| Plumbene |  | ~200 |  |  | +1250 | [111] |
| Arsenene (ultra-flat) | SiC(0001) | ~200 |  |  | Approx. mid gap | [109] |
| Antimonene (ultra-flat) | SiC(0001) | 300 – 350 |  |  | Approx. mid gap | [109, 277] |
| Bismuth |  |  |  |  |  |  |
| (111) bilayer |  | ~200, ~600 |  |  | Approx. mid gap | [84, 86, 87] |
| (110) bilayer |  | 88.9 |  |  | Approx. mid gap | [103] |
| (ultra-flat) | SiC(0001) | 506 – 560 | SiC(0001) | 800 | Approx. mid gap | [52, 109, 277] |
| (ultra-flat) | Si(111) | 800 |  |  | Approx. mid gap | [278] |
| (ultra-flat) |  | Trivial |  |  | Approx. mid gap | [109, 277] |
| Ultra-thin Dirac Semimetals ||||||||
| Na$_3$Bi |  | 310 | Si(111) | >300 | Approx. mid gap | [53, 178] |
| 1T'-WTe$_2$ |  | 0, >100 | HOPG | 55, 62 | Approx. mid gap | [44, 157, 161, 169] |
| 1T'-WSe$_2$ |  | 36, 116 | HOPG | ~120, 129 | -130 | [44, 166, 167] |
| Room temperature ($T$ = 298 K): 3.53 $k_BT \approx$ 90.6 meV ||||||||

**Table 1. Atomically thin Quantum Spin Hall Materials**

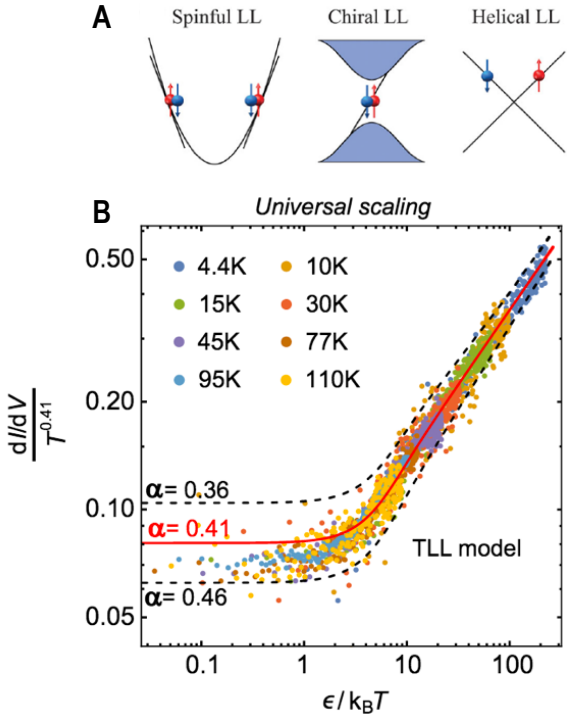

The resulting size of the spin-orbit gap in any real QSH material is thus a combination of effects arising from the lattice structure itself, the atomic number of the constituent species, and additional contributions from interactions with substrate or adsorbates. To date, several QSH materials have been predicted and experimentally confirmed, and are enumerated in Table 1.

**Figure 3. Signatures of the Tomonaga Luttinger edge liquid.** (A) Schematic band dispersion of three different 1D electronic systems that can host Tomonaga-Luttinger liquids (TLL), including a conventional (spinful) 1D metal with parabolic dispersion (left), a chiral edge state with linear dispersion as in a quantum Hall state (middle), and a helical system representing a QSH insulator (right). Reproduced with permission.[21] Copyright 2015, APS. (B) Universal scaling of the helical edge state LDOS of the QSH insulator bismuthene, probed via scanning tunneling spectroscopy, demonstrating the presence of a TLL. Reproduced with permission.[74] Copyright 2020, Springer Nature.

## 4. Interactions on the Edge: The Helical Tomonaga-Luttinger Liquid

Electron correlations are known to be strongly enhanced in low-dimensional electronic systems.[71] In strictly 1D, for instance the helical edges of a QSH insulator,[26, 27, 72] the Fermi surface is composed of only two Fermi points $\pm k_F$ for forward and backward propagating electrons. The reduced dimensionality has the effect that electronic excitations become strongly correlated even at minimal interaction strength resulting in a breakdown of the Fermi liquid picture [73]. Rather, such interacting 1D electron liquid is predicted to have a so-called Tomonaga-Luttinger Liquid (TLL) ground state, distinguishable from a Fermi liquid chiefly by spin and charge separation as well as the formation of spin and charge density waves,[73] with plasmons and spin density waves being the fundamental low-energy excitations.

Interactions in a TLL are parameterized by Luttinger parameter $K$, which characterizes the strength of the interactions. At low energies, $K$ can be represented[74, 75] as

$$K = \left(\frac{1+y_2-y_4}{1+y_2+y_4}\right)^{\frac{1}{2}} \tag{7}$$

where $y_i = g_i (2\pi\hbar v_F)^{-1}$, $g_2(g_4)$ denote the forward scattering amplitudes and the same (opposite) Fermi points, and $\hbar v_F$ is the Fermi velocity. For the experimentally practical case of Coulomb interactions in 1D wire screened by a conducting plane, one can let

$$g_2 = g_4 = V(q=0) = \int V(x)e^{-iqx}dx\Big|_{q=0} \tag{8}$$

where $V(q = 0)$ is the Coulomb interaction between neighboring electrons in the long-wavelength ($q = 0$) limit. Then, $K$ can be approximated as[31, 76]

$$K \approx \left(1 + \frac{V(q=0)}{\pi\hbar v_F}\right)^{-\frac{1}{2}}. \tag{9}$$

Luttinger parameter $K = 1$ describes a noninteracting system, whereas $0 < K < 1$ and $K > 1$ correspond to repulsive and attractive interactions, respectively.

Interactions strongly affect the tunneling conductance $G$ of electrons into a TLL, which reduces from the quantized value $G_0 = 2e^2h^{-1}$ (ballistic limit) by a characteristic power law dependence, simultaneously on temperature[31] at low tunneling bias ($eV \ll k_BT$), $G \propto T^\alpha$, and tunneling energy $eV \gg k_BT$, $\frac{dI}{dV} = V^\alpha$, with the same universal power law exponent $\alpha$.

The universal exponent $\alpha$ depends on the number of 1D channels[77] and whether electrons tunnel into the TLL bulk or only a single end.[78] More fundamentally, is related to the Luttinger parameter $K$ as $\alpha = C(K + K^{-1} - 2)$. Depending on the class of 2D TI, spin degeneracy may further be broken, distinguishing spinful ($C = ¼$, quantum Hall) and helical ($C = ½$, quantum spin Hall) TLLs (**Figure 3A)**. For a helical TLL at the QSH edges[79] $C = ½$ due to tunneling of excitations with fractional charge $e/2$ between energy minima in the helical edge.[31] Chiral TLLs (Figure 3A) also exist, which possess a single, unidirectional, spin-degenerate edge mode and are found in quantum anomalous Hall insulators and fractional quantum hall insulators,

Such electronic correlations in a helical TLL can be resolved in electron transport, or by local probe spectroscopy such as scanning tunneling microscopy/spectroscopy (STM/STS), and can offer unprecedented insight into the electronic structure. Especially STM/STS offers direct access the tunneling differential conductance in real-space, which is discussed in detail in Section 6.3.

For an individual tunnel junction with a TLL, the temperature and bias dependencies of the differential tunneling conductance may be expressed in a single expression[74, 78, 80, 81] as

$$\frac{dI}{dV} = AT^\alpha \cosh\left(\frac{eV}{2k_BT}\right)\left|\Gamma\left(\frac{1+\alpha}{2} + i\frac{eV}{2\pi k_BT}\right)\right|^2 \qquad (10)$$

where A is a constant and $\Gamma(x)$ is the gamma function. The above form describes a sharp suppression of the measured local density if states at $\varepsilon = 0$ eV that can be directly probed as a function of STM junction temperature and STM bias $V$ (see Section 6.3.2). More importantly, by plotting the measured differential conductance, normalized by the $T^\alpha$, in dimensionless energy units $(E-E_F)(k_BT)^{-1}$, a universal scaling plot can be achieved in which data points measured at different temperatures collapse onto a single trace. Such universal scaling is often regarding as a smoking gun signature of a TLL state.[74, 78]

## 5. Atomically thin QSH Materials

Since the discovery of graphene as the first predicted atomically thin quantum spin Hall material, considerable effort has been devoted into isolating and synthesizing atomically thin materials with related crystal structures[43, 82] and larger spin-orbit induced topological gaps. Indeed, material science now has access to a library of layered materials[43, 52, 53, 83] with lattices of hexagonal symmetry that provide a vast pool for potential discovery of the QSH state.

### 5.1. Xenes

The physical reason for graphene's small topological gap is its small atomic number combined with the presence of inversion symmetry of its lattice, both giving rise to weak spin orbit coupling. The former can be mitigated by considering honeycomb lattices composed of group IVA and group VA elements of higher atomic number, while inversion symmetry can be broken in certain buckled or puckered hexagonal lattices. Furthermore, the QSH gap may be modified by strain, electric and magnetic fields, and the chemical environment of the material. As a result, the Xenes show the widest range of intrinsic bulk topological gaps amongst all known QSH materials, ranging from likely ~42 μeV measured[51] in graphene (consistent with predictions[48-50, 65, 66]) up to ~800 meV measured in ultra-flat bismuthene.[52]

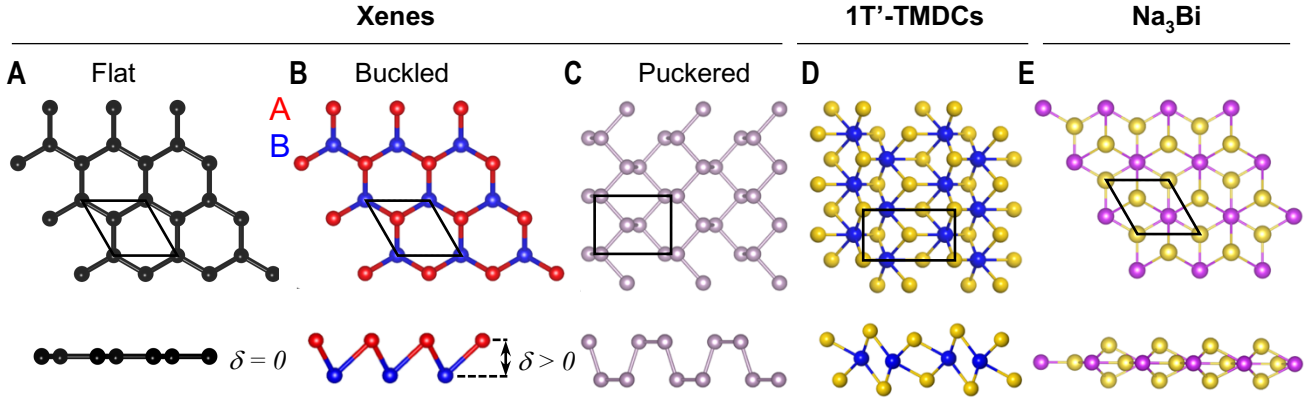

**Figure 4. Atomic structure of different QSH materials.** (Top) Basal plane and (bottom) side views of QSH lattices. (A-C) Xene atoms coordinate with 3 nearest neighbors and adopt primarily three structural polymorphs: (A) flat, honeycomb lattice, (B) buckled, wherein the inequivalent sublattices A and B (red and blue) are offset vertically by distance $\delta > 0$, and (C) puckered, which has characteristic armchair ridges in the side view. (D) In 1T'-phase transition metal dichalcogenides, transition metal atoms (blue) dimerize along one direction, and each one coordinates with six chalcogen atoms (yellow) (E). In a single layer of $Na_3Bi$, Bi atoms (violet) form a hexagonal sublattice wherein each Bi atom coordinates with 9 Na atoms (yellow), resulting in 3 planar sublayers (i) Na, (ii) Na/Bi, (iii) Na.

Atomically thin honeycomb materials composed of single-element group IVA and VA species have been predicted to host a variety of topological states tunable by strain engineering and chemical functionalization.[15, 36, 43, 66, 84-109] Aside from atomically flat Xenes (**Figure 4A**), such as graphene itself, two basic structural isomorphs of Xenes have been reported: (1) a buckled structure, shown in Figure 4B, where the two interpenetrating hexagonal sublattices A and B are no longer equivalent and are vertically offset by buckling parameter $\delta > 0$,[43] and (2) a puckered structure, shown in Figure 4C, which features armchair ridges in the side view. These three isomorphs are governed by distinctly different underlying physics, as detailed below.

*5.1.1. Graphene and Group IVA Buckled Honeycomb Lattices*

In ideal graphene, carbon atoms are arranged in a planar ($\delta = 0$) honeycomb structure, as shown in Figure 4A. In addition, there may be vertical buckling of the lattice ($\delta > 0$, see Figure 4B), leading to an inequivalence of the A and B sublattices. Its atomic orbitals are *sp*$^2$ hybridized with nearest neighbors strongly bonded by σ-σ bonds. Mobile electrons can hop to nearest neighbors via the lattice-commensurate, unsaturated $p_z$ orbitals that hybridize to form a π band network. The low-energy physics of graphene is dominated by six inequivalent band crossings at the K (K') points located at the corners of the hexagonal Brillouin zone, which exhibit a linear (Dirac) dispersion at low energy. The Kane-Mele model predicts SO gap at these Dirac nodes in the 2D bulk thus realizing a quantum spin Hall state.[15] However, due to its planar structure, the spin-orbit coupling in graphene only occurs via inherently weak processes, such as virtual transitions to σ- and *d*-orbitals.[15, 48-50, 65, 66] SOC in graphene is unlikely to open a spin-orbit gap larger than $E_g \sim 50$ μeV.[48-50, 65, 66] While there have been predictions[88] and reports[93, 110] of much stronger induced SOC in graphene, no definitive measurement of graphene's intrinsic SO gap has been reported to date despite its experimental isolation over 15 years ago.

As argued above, for the simplest case of atomic spin-orbit coupling, the SOC strength is often considered to increase as the fourth power of the atomic number ($\sim Z^4$). This would indicate that honeycomb lattices composed of larger *Z* atomic species may feature significantly larger SOC gaps. Indeed, a clear $Z^4$ trend appears to be supported by density functional theory calculations[49, [50, 106, 111] of group IVA honeycomb lattices (see Figure 2A). All freestanding, low-buckled group IVA Xenes with $Z > 6$ are predicted to possess an intrinsic spin-orbit gap that is measurable at reasonable experimental temperatures.[84, 85, 106, 112] Freestanding stanene (Sn, $Z = 50$), for example, is predicted to possess an intrinsic SO gap of ~ 73-100 meV.[90, 106, 112]

Additional SOC contributions may arise from buckling of the Xene lattice, effectively increasing the overall SOC strength. The $Z > 6$ (Si, Ge, Sn, and Pb) elements are predicted to stabilize in "low-buckled" ($\delta > 0$) structures, even in their relaxed, freestanding forms.[85, 90, 106, 112-115] The larger interatomic bond lengths in these materials weaken the π-π orbital overlap, which destabilizes the planar structure and causes the material to become slightly buckled.[85] Buckling allows σ and π bands to mix, giving rise to a slight sp$^3$ character of the bonds.[85] While the intrinsic band structures of all higher-$Z$ Xenes are predicted to be similar to that of graphene,[85, 90, 106, 112-115] the increased orbital overlap due to buckling contributes to a slight increase in the SO gap.[85]

As a consequence of their orbital structure, QSH behavior in Xenes may be tuned by orbital hybridization (for example, via substrate interactions[116, 117]), or induced by lattice strain,[105] as well as by chemical functionalization with impurities or functional groups.[90, 118] In buckled stanene, for example, compressive strain leads to an increase of the $p_{x,y}^+$ orbital energy above the Fermi level at Γ, rendering the material metallic.[105, 106, 119] Furthermore, hybridization may either saturate $p_z$-orbitals or break A-B sublattice symmetry, moving the π-bands far away from the Fermi level to open a gap at K(K').[101] For the case of chemical functionalization, $p_z$ orbitals may become saturated, moving the spin-orbit gap at K(K') to far from the Fermi level, in which case the low-energy physics becomes dominated by a topologically nontrivial gap of ~250-300 meV at Γ.[90] Chemical functionalization of stanene by fluorine, for example, may saturate the $p_z$ orbitals and shift them away from the Fermi energy.[90] In contrast, stanene functionalized by hydrogen (*stanane*), is expected to possess a trivial bandgap as it should not have the necessary *s-p* band inversion to support a QSH state.[90, 118]

Realizations and predicted topological gaps of group IVA Xenes are summarized in Table 1. For additional experimental details regarding lower-*Z* Xenes, the reader is directed to an excellent review by Molle et al.[43]

*5.1.2. Group VA Puckered and Buckled Honeycomb Lattices*

Aside from the group IVA elements, also the group VA elements like phosphorous,[120, 121] arsenic,[122] antimony,[97, 105, 123, 124] and bismuth[84, 89, 125-129] can form honeycomb lattices. Usually, they adopt stable puckered and buckled lattices, as illustrated in **Error! Reference source not found.**B,C. Individual units of group VA buckled and puckered 2D honeycombs are typically referred to as "bilayers," which is a nomenclature reflective of their bulk crystal structure wherein each periodically repeatable unit in vertical direction is referred to as a "layer". Importantly, group VA elements contain one additional electron compared to group IVA elements, giving them 5 valence electrons in total which saturate the valence molecular orbital,[37] as three electrons participate in bonds, while the remaining two electrons can then saturate the remaining molecular valence orbital. As such, most group VA honeycombs do not exhibit the same Dirac physics in their 2D bulk that governs group IVA honeycombs. Instead, conductance and valence bands are usually separated in unstrained lattices by a trivial band gap between $p_{x,y}$ and $p_z$ orbitals.[37] A prominent exception is the case of bismuth bilayers, whose strong SOC inverts $p_{x,y}$ and $p_z$ orbitals of opposite parities[130] and is topologically nontrivial.

Phosphorous ($Z$ = 15) and arsenic ($Z$ = 33) honeycombs have been predicted[120, 122] to adopt two different structural isomorphs; a buckled structure, similar to the group IVA Xenes wherein the two hexagonal sublattices are offset by $\delta > 0$, as well as a puckered structure, which black phosphorous[131] and black arsenic[132] can both adopt. As predicted by DFT, both buckled arsenic honeycomb monolayers (*arsenene*) and stacked black phosphorous bilayers (*phosphorene*) are expected to undergo a QSH phase transition precipitated by strain-induced band inversion of *p*-orbitals,[121, 133] with a gap up to $E_g$ = 43 meV for arsenene[133] and $E_g$ = 92.5 meV for stacked phosphorene.[121]

Buckled antimony (Sb) ($Z$ = 51), or Sb(111), bilayers, are expected to exhibit a rich variety of thickness-dependent electronic phases.[123] Predicted to be trivially gapped for thicknesses between 1-3 bilayers, this material is expected to undergo a thickness-induced topological phase transition at 4-7 bilayers.[123] Increasing the material's thickness further, from 8 bilayers up to 22 bilayers (~7.8 nm), it is predicted to result in a 3D topologically insulating state. Films above this thickness threshold are predicted to be a topological semimetal.[123]

The electronic structure of bismuth ($Z$ = 83) surfaces have been extensively investigated, both theoretically[84, 86, 87, 89, 97, 103, 125, 130, 134, 135] and experimentally[59, 126-129, 136-147]. Recently, bismuth has been shown to be a higher order topological insulator (HOTI),[129] possessing helical "hinge" states that consist of one-dimensional gapless Kramers pairs, like the QSH edge, around specific facets of the crystal, which exists even for 3D bulk samples. As a consequence, both monolayers of Bi(111) (buckled honeycomb structure, **Error! Reference source not found.**B) and Bi(110) (puckered honeycomb structure, **Error! Reference source not found.**C), have been predicted to host gapless edge modes, consistent with those in a quantum spin Hall insulator, tunable by strain and electric field.[84, 86, 87, 103] Different from the QSH states that host topologically nontrivial states on all edges, however, higher order topology in Bi(111) gives rise to *hinge* states only on every other zigzag edge of the hexagonal crystal facets.[129] The SO gap of a single, relaxed Bi(111) bilayer is predicted to be ~200-600 meV,[84, 86, 87] whereas that of a single, relaxed Bi(110) bilayer is predicted to be $E_g$ ~ 90 meV.[103]

*5.1.3. Group IVA and VA Multi-Orbital Honeycomb Lattices*

Monolayer Xene phases having a molecular orbital structure with low-energy physics dominated by a network of $p_x$ and $p_y$ orbitals, rather than a $p_z$ orbital network like in group IVA Xenes, have been predicted to offer an attractive paradigm for large-gap QSH insulators,[36, 95, 97, 109, 148] with predicted band gaps up to ~1 eV.[148] Similar to the π orbital network in freestanding group IVA Xenes, the multiorbital $p_{x,y}$ network has six Dirac nodes at the K(K') points in the absence of SOC.[52, 109] However, strong on-site atomic SO coupling, a first-order interaction that is much stronger than, for example, the second-order, next-nearest neighbor SO interaction in freestanding group IVA Xenes[36, 52, 95, 109, 148] results in huge intrinsic QSH gaps.

In these materials, the $p_z$ orbitals are chemically saturated via bonds with an atomic-lattice-matched substrate or by attached functional groups to the Xene atoms such that each atom sees the same bonding environment, which "filters" the orbital away from $E_F$. These Xenes are ultra-flat ($\delta = 0$),[36, 52, 89, 97, 107, 149] and hence usually possess larger lattice constants compared to their buckled ($\delta > 0$) counterparts,[52, 140, 149, 150]. Such ultraflat Xenes have already been realized with several large $Z$, group IVA and VA elements including tin ($Z = 50$), antimony ($Z = 51$), lead (Pb = 82), and bismuth ($Z = 83$).[52, 107, 149, 151, 152] Ultra-flat bismuth on silicon carbide (0001) substrate producing a measured band gap of ~800 meV,[52] the largest intrinsic gap of any QSH insulator thus far.

**5.2. Ultra-Thin Weyl and Dirac Semimetals**

The quantum spin Hall state may also occur in two-dimensional layered crystals with non-honeycomb lattices, or in topologically non-trivial bulk semimetals, such as 3D Dirac[153, 154] or Weyl[44] semimetals, in which quantum size effects can lead to a topological phase transition when the material is thinned down to a single (or a few) layers.

*5.2.1. 1T'-phase Transition Metal Dichalcogenides*

Transition metal dichalcogenides (TMDs) are a family of layered van-der-Waals materials with $MX_2$ stoichiometry, where M = transition metal and X = chalcogen, typically S, Se, or Te. Each monolayer is composed of an atomic M layer, sandwiched between X layers, whereby several structural polymorphs exist depending on M coordination symmetry and layer stacking.[155] The different structural polymorphs, in combination with different combinations of transition metal group and chalcogen species, gives rise to a rich pool of different electronic structures, including large bandgap semiconductors, metals, and semimetals. In particular, the group-VI TMDCs (Mo, W) combined with chalcogens S, Se, and Te are predicted to be rich with topologically non-trivial phases when stabilized in their 1T' structural polymorph.[44, 156-160]

The 1T' phase is characterized by orthorhombic crystal structure belonging to the *Pmn21* space group, as shown in **Error! Reference source not found.**D. The 1T' phase is similar to the 1T (tetragonal) phase but distorts to form dimerized transition metal chains within each atomic layer resulting in a uniaxial doubling of the crystal unit cell. This lattice distortion inverts the chalcogen *p* and transition metal *d* orbitals, giving rise to a Dirac dispersion at the Fermi level in the absence of SOC.[44] In the 3D bulk forms of 1T'-TMDs, type-II Weyl physics,[156, 160] loop nodal band crossings,[158, 159] and higher order topological properties[158, 159] have all been predicted for the Te compounds.

The quantum spin Hall state was first predicted by Qian et al. in monolayers of 1T'-$MX_2$ (M = W, Mo and X = S, Se, Te) using DFT,[44] with strain and electric-field tunable spin-orbit gaps ~0.1eV in magnitude. The electric field tunability[44] is facilitated via the Stark effect as transition metal and chalcogen sublattices are spatially separated in z-direction. A transverse electric field can thus induce a topological phase transition by breaking inversion symmetry and inducing a Rashba splitting of degenerate bands near $E_g$.[44]

The quantum spin Hall state in monolayer $WTe_2$ has since been confirmed experimentally,[54, 55, 161-163] where a QSH gap of ~55 meV was discovered, and stable QSH transport signatures were observed up to *T* = 100 K.[54, 55] Despite overwhelming experimental evidence of the QSH state,[54, 55, 161, 164-169] several interesting questions remain, in particular with regard to the exact size and nature of the topological gap and its susceptibility to doping, local strain, and electric fields. DFT reports seem to disagree about the size of the QSH gap. Qian's initial prediction reports a negative band gap in freestanding $WTe_2$,[44] while more recent calculations report a band gap up to 141 meV.[157] Several experimental studies have further revealed a small but finite density of states at the Fermi level within the 2D bulk.[161, 162, 164-168] Yet, stable QSH signatures with little bulk conduction have been found in electron transport experiments.[54, 55] The contrasting experimental evidence may highlight the sensitivity of the topological band gap to strain, electric fields, and doping in different substrate environments.[163, 169]

An interesting aspect of 1T'-$MX_2$ is the role of strong electron correlations in 2D, possibly stabilizing the QSH phase. Several reports have claimed a weak pseudo-gap suppression in the density of states[162, 164, 166, 170] when the material is doped such that the Fermi level comes to lie in the conduction band.[162, 166] This pseudo-gap in this case is distinct from the QSH gap, and remains strictly pinned at the Fermi level, regardless of doping level.[162] Such pseudo-gap has been argued to result from Coulomb correlations of carriers at the Fermi level, opening an Coulomb gap[162] in the density of states. As both electron and hole pockets are present close to the Fermi energy,[44] such Coulomb gap would likely be of excitonic nature as first pointed out by Song et al.[162]

Although the largest volume of theoretical and experimental work published concerns 1T'-$WTe_2$, other TMDC chemical compositions may offer further advantages. $WTe_2$ is thermodynamically stable in the 1T' polymorph, whilst other TMDs have been shown to be bi-stable. $WSe_2$ crystals, for instance, have been shown to exhibit coexisting crystalline domains of 1T' and 1H lattice structure,[166, 167, 171] the latter being a trivial insulator with a large electronic bandgap (1.9 eV).[166] This offers the tantalizing possibility to engineer[166] ordered and atomically abrupt 1T'- 2H crystal phase boundaries, whose interface can host one-dimensional conducting boundary modes.[166] Preliminary phase boundary engineering has already been demonstrated in 1T'/1H WSe$_2$[171] and in other MX$_2$ materials.[172] A recent review by Li et al.[173] provides a comprehensive summary of 1T'-MX$_2$ quantum spin Hall properties.

*5.2.2. Na$_3$Bi and Cd$_3$As$_2$*

Other 3D topological semimetals including Cd$_3$As$_2$ and Na$_3$Bi, have been predicted to undergo a thickness-dependent topological phase transition to a QSH insulator in the limit of a single or a few atomic layers.[153, 154] Of these two materials, Na$_3$Bi[174, 175] has been successfully synthesized and characterized at the ultra-thin limit.[53, 176] The physics of the topological phase transition is similar in both materials.

The crystal structure of monolayer Na$_3$Bi has P63/mmc symmetry as shown in Figure 4E. Bulk Na$_3$Bi consists of Na-Na/Bi-Na tri-layers wherein each tri-layer has a middle layer of interpenetrating hexagonal Na and Bi sublattices sandwiched between hexagonal layers of Na, and adjacent tri-layers are rotated by 60 with respect to each other.[175] This material hosts 3D Dirac fermions owing to a linear electronic dispersion in all momentum directions.[175] Na$_3$Bi has been shown theoretically to undergo a topological phase transition to a QSH insulator[177] at a thickness below 7 layers due to quantum confinement. Monolayer and bilayer Na$_3$Bi have furthermore been shown to be electric field-tunable,[53] and monolayers are predicted to exhibit spin-valley polarization.[178]

In the absence of SOC, monolayer Na$_3$Bi is predicted to be trivially gapped at Γ, with Na and Bi $s$ orbitals forming the conduction band minimum (CBM) and Bi $p_{x,y}$ orbitals forming the valence band maximum (VBM).[178] However, SOC drives $s$-$p$ band inversion at Γ such that the VBM gains $s$ orbital character and the CBM gains $p_{x,y}$ orbital character,[178] giving rise to gapless edge modes in the monolayer and an SO band gap within the 2D bulk, which has been measured to be as large as $E_g \approx 360$ meV,[53] consistent with the predicted SO gap of $E_g = 310$ meV.[178] Similar QSH properties have also been predicted and measured in bilayers,[53, 179] with a measured quasiparticle gap of about $E_g = 300$ meV.[53, 179]

## 6. Experimental Methods to Investigate the QSH State

The hallmark signatures of the QSH state in any material are the concurrence of a gapped 2D bulk band structure and a finite 1D density of states at the edge. Both features are experimentally accessible in a variety of experimental techniques, including optical, local-probe and transport spectroscopies, as well as their combinations.[52, 53, 107, 129, 161, 166, 167] Usually, only a combination of the experimental techniques discussed, and the particular QSH signatures they provide, can be regarded as a comprehensive confirmation of the QSH state in a particular material.

### 6.1. Electron Transport

The first experimental confirmation of the quantum spin Hall state was demonstrated using electron transport spectroscopy in HgTe/CdTe[16, 24, 180] and later also in InAs/GaSb[22, 23] heterostructures. Shortly after the initial theoretical prediction of the QSH state in HgTe/CdTe semiconductor heterostructures,[24] König et al.[17] confirmed the QSH state in Hall bar measurements of high-mobility 2D quantum wells with topological band inversion (**Figure 5A**).

Being able to tune the electrochemical potential from *n*-type to *p*-type across a small energy gap, they were able to detect a small band gap through a suppression of the Hall bar's longitudinal conductance was observed throughout the band gap in the topological regime. Such residual conductance, independent of the Hall bar width, thus indicated that current was confined to the samples' boundaries. Perfect ballistic conduction with a quantized conductance of $G_0 = 2\ e^2\ h^{-1}$ – consistent with two forward propagating modes, one each per edge – was observed for very short channel lengths (~1 μm), comparable to the elastic electron mean free path,[17] and provided initial indications of the presence of helical edge modes. Furthermore, a strong suppression of the residual conductance upon application of a magnetic field was found consistent with time-reversal symmetry breaking and the opening of a Zeeman gap at the Kramers' degenerate points in the edge. König et al.'s magnetoconductance measurements are shown in **Figure 6A**. All these signatures combined were strong initial evidence for the presence of a QSH state for the first time in a real material. The QSH state in HgTe/CdTe has since been further confirmed by non-local electrical measurements[18] (see discussion below), as well as by measurements in HgTe/CdTe quantum point contacts (QPCs) in which opposite helical edges were found to interact by inter-edge tunneling and hybridization.[181]

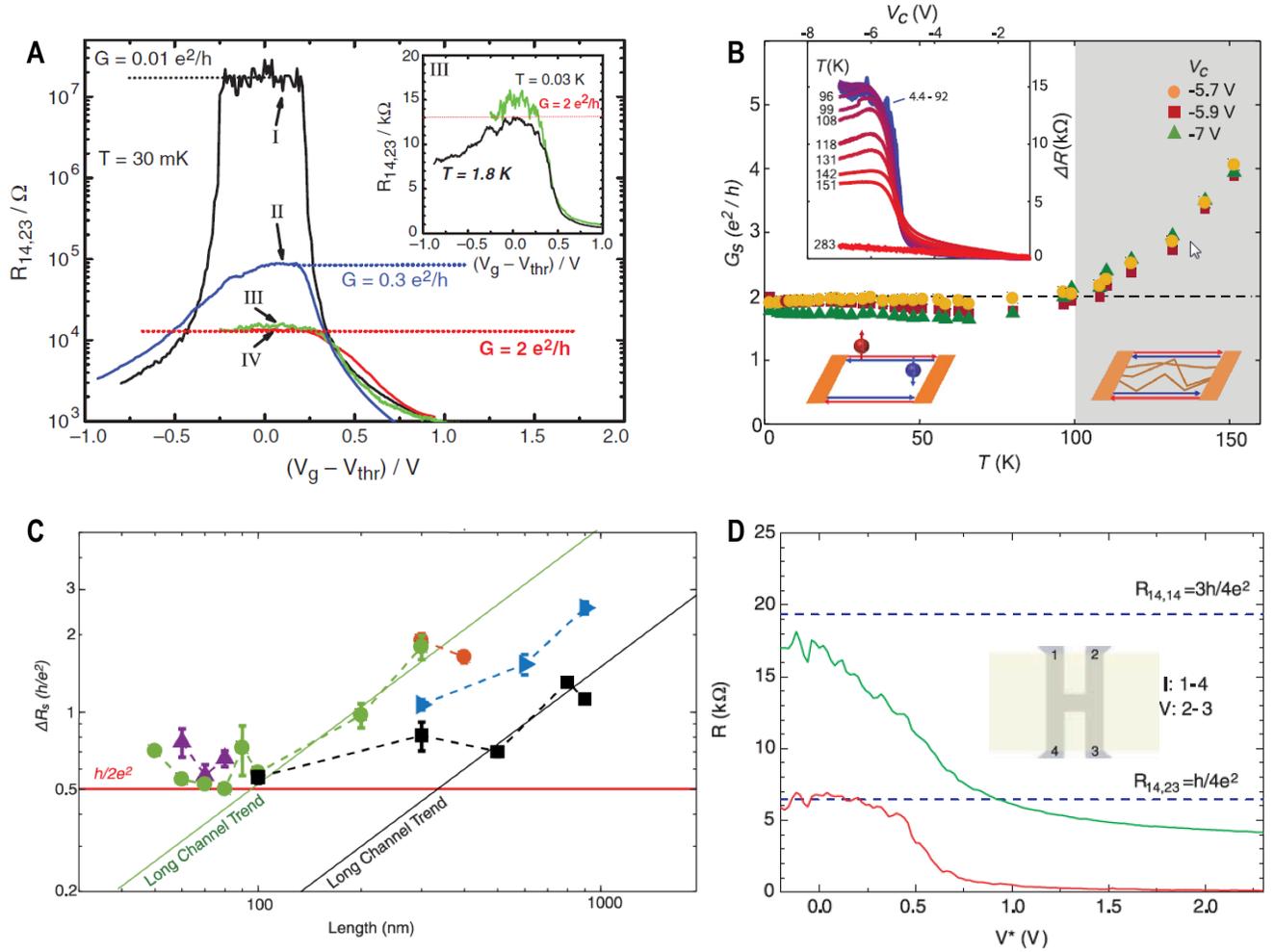

**Figure 5. Electron transport spectroscopy of QSH insulators.** (A) Conductance as a function of gate voltage for four different HgTe/CdTe semiconductor heterostructure Hall bar devices. A maximum of finite resistance in the longitudinal Hall bar resistance over a given gate voltage range indicates the QSH gap, a finite resistance within the gap indicates edge state conduction. The different traces show different Hall bar geometries with length L, width W, and thickness d. For reference, I: d = 5.5 nm; II, III, IV: d = 7.3 nm, and I, II: L x W = 20 x 13 µm$^2$, III: L x W = 1 x 1 µm$^2$, and IV: 1 x 0.5 µm$^2$. In short Hall bar devices (III, IV), quantized edge state conductance with $G_0 = 2$ e$^2$ h$^{-1}$ is realized, indicating ballistic transport. Inset: Device III at two different temperatures as indicated. Reproduced with permission.[17] Copyright 2007, AAAS. (B) Temperature-dependent conductance of a 100 nm channel in a 1T'-WTe$_2$ device [54], showing quantized conductance up to $T \sim 100$ K. (Inset) Temperature dependence of the channel resistance as a function of back gate voltage $V_C$. (C) Length dependence of edge state resistance for five 1T'-WTe$_2$ devices [54], at $T = 4$ K, showing a linear dependence of the resistance on length, for devices L > 100 nm. (B,C) Reproduced with permission.[54] Copyright 2018, AAAS. (D) Non-local transport in HgTe quantum well QSH insulators, patterned into H-bar devices (Inset). The two- and four-terminal resistances at $T = 1.8$ K show a non-local resistance. Expected values based on the Landauer-Buttiker formalism are indicated by dashed lines. Reproduced with permission.[18] Copyright 2009, AAAS.

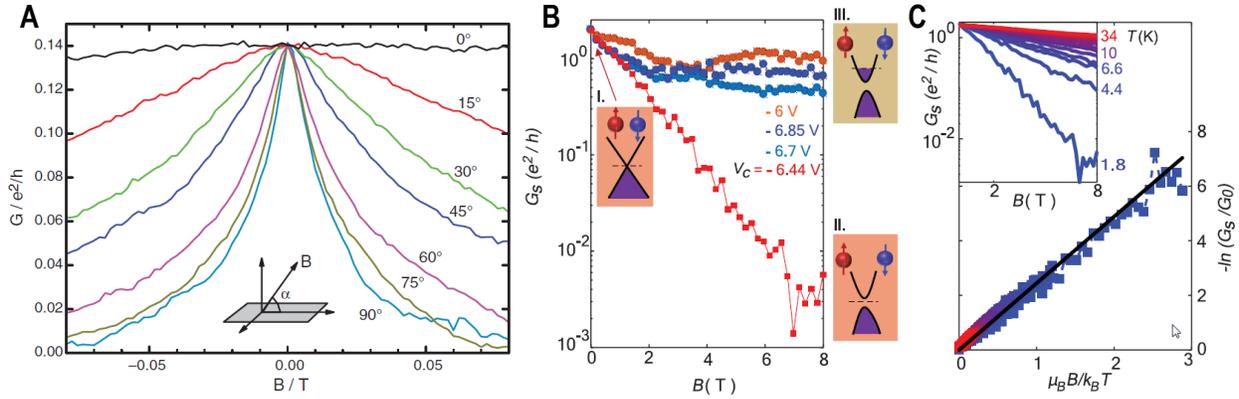

**Figure 6. Time-reversal symmetry breaking by a magnetic field.** (A) Negative magnetoconductance at $T = 1.4$ K of an inverted HgTe/CdTe heterostructure device showing how a different magnetic field tilt angles suppress the helical edge state conduction. Reproduced with permission.[17] Copyright 2007, AAAS. (B) Similar data for a 1T'-WTe$_2$ device ($T = 1.8$ K), for four different values of a back-gate voltage ($V_C$) (see inserts). An exponential suppression of the edge state conduction is only seen when the Fermi level is tuned to within a Zeeman gap in the edge state dispersion. (C) Collapsing the normalized magnetoconductance measured at different temperatures onto a single curve allows to extract the edge states' electron $g$-factor (B, C from [54]). Reproduced with permission.[54] Copyright 2018, AAAS.

Ever since the first demonstrations by König et al., transport signatures as described remain the benchmark for any QSH claim, and have since also been applied to atomically thin QSH materials.[54, 55, 182] As for any atomically thin material, the key-challenges in transport measurements of QSH insulators are usually nanofabrication related as materials are often temperature unstable or air-sensitive. This imposes some constraint on nanofabrication techniques employed and challenges the formation of low-Ohmic contacts to a highly confined QSH edge. Fei et al.[55] first demonstrated QSH transport signatures in $h$-BN encapsulated monolayer 1T'-WTe$_2$, in which they observed a temperature independent residual conductance at $T < 10$ K, with thermally activated conduction between $T = 10 - 100$ K. Wu et al.[54] further confirmed this, reporting an even weaker temperature dependence below $T < 100$ K, as shown in Figure 5B. This seemed to indicate a stable QSH state with a comparatively large topological gap, for the first time promising QSH based applications above liquid helium temperatures.

Strong evidence for the QSH state in WTe$_2$ was obtained from magnetoconductance measurements,[54, 55] allowing an accurate extraction of the spin-orbit enhanced electron $g$-factor. Both Fei et al.[55] and Wu et al.[54] observed thermally activated conductance (Figure 6B), across a small, magnetic field induced Zeeman gap at the Kramers degenerate (Dirac) points. From logarithmic scaling of the conductance[54] (Figure 6C) a $g$-factor of $g = 4.8$[54] and $g = 7.5$[55] were found by Wu et al.[54] and Fei et al.,[55] respectively.

A key advantage of atomically thin QSH materials is their versatility with regard to contact geometries. The atomically thin nature of the material lifts the constraint of edge contact to, for example, a semiconductor mesa structure.[17] This had not only allowed to align a large number of electrodes with different probe spacing along the QSH edges to investigate the robustness of edge conduction against scattering,[54] but also to align contacts on top of the 2D surface, facilitating new and more complex contact geometries to e.g. disentangle contributions from bulk and edge to the total conductance.[55] Using such multi-probe geometries, Fei et al. were able to demonstrate that bulk conduction is negligible in $WTe_2$ below ~10K, while Wu et al. was able to obtain a length dependence of the conductance for different probe spacing (**Error! Reference source not found.**C).

As a key experiment to prove the QSH state have been non-local electrical measurements.[18] In transport measurements of conventional conductors, such as e.g. a conventional 2D electron gas, the current versus voltage relationships would follow Ohm's law in the linear response regime. This implies that current follows the distribution and direction of the applied electric field, and a voltage drop is only detected (*local voltage*) along the current path between the current injecting electrodes. In topological insulators – such as the quantum spin Hall insulator – these basic electrostatic concepts are fundamentally changed as the QSH bulk is insulating and current flow is confined to the QSH boundaries regardless of the direction of the electric field. Due to the helical nature of the edge, with current flow around the entire sample boundaries, a measurable voltage drop may arise far from (and not necessarily in between) the current-injecting leads (*non-local voltage*). Indeed, the strictly one-dimensional nature of the QSH helical edges allows for a facile quantitative analysis of the non-local voltages in the framework of the Landauer-Büttiker formalism, as first shown by Roth et al. in HgTe/CdTe heterostructures[18] (**Error! Reference source not found.**D). More recently, similar non-local transport signatures have also been demonstrated in atomically thin QSH insulators such as $WTe_2$[55] and $Na_3Bi$.[183]

To date, electron transport spectroscopies have provided the most direct access to electronic properties of 2D topological systems. However, they are also inherently limited to spatially averaged electronic information and information about the helical edge can only be obtained indirectly – by evaluating the magnitude of the quantized residual conductance. However, the challenges concomitant with transport measurements, such as the required nanofabrication processes, have so far limited transport experiments to air-stable or encapsulated materials, or have restricted measurements to within vacuum environments.[184, 185]

Another drawback of transport measurements is that certain electronic information, such as the confinement length of the helical edge to the sample's boundaries, local variations in the disorder potential and electrochemical potential, and the microscopic origin of scattering remain inaccessible as they require techniques with real-space information.

A key remaining open question for any QSH materials is the strong length dependence[17] of the quantized conductance at the QSH edge obtained in transport experiments. Usually, rather large edge state resistances[183] are observed unless very short channels[17, 54] are measured with lengths of the order of the 2D mean free path. From basic considerations, one should expect a strong suppression of backscattering from non-magnetic impurities at the helical edge and hence ballistic conduction over considerable length scales. This has however not been observed in any QSH material so far and raises questions about the coveted topological protection. The abundance of scattering observed has since triggered a substantial amount of theoretical investigation into the nature of back-scattering mechanisms[31, 186-189], discussing dilute magnetic impurities [26, 27, 31, 186-188] or metallic puddles carrying magnetic moments,[26, 27, 189] as well as inelastic backscattering as possible mechanisms.

**6.2. Optical Spectroscopy**

Perhaps the most direct way to infer electronic structure information in solids is to measure the electronic band dispersion and Fermi surface by angle-resolved photoemission spectroscopy (APRES). APRES measures the energy and momentum distribution of photoemitted valence electrons from a material's surface, following illumination, thus being able to directly 'visualize' the band structure. Due to strong advancement in light-source, detector, and cryogenics technologies, samples can be measured at temperatures down to ~4.2 K, with energy and momentum resolution better than 1 meV[190, 191] and 0.0006 Å$^{-1}$,[192] respectively.

Furthermore, ARPES is an extremely surface sensitive technique (typically 10 nm at incident photon energy $E$ = 10 eV[193]), an attribute which has made ARPES the key analysis tool to resolve the surface band structures of 3D topological insulators.[34, 194] Ironically, for the same surface sensitivity, ARPES measurements of the bulk band structure and bulk gaps in 2D topological insulators have only recently appeared, hand in hand with the emergence of atomically thin QSH materials. In particular, the authors are not aware of any reports on ARPES measurements in semiconductor heterostructure based QSH insulators. As possible reasons, we suspect that the active layer is covered by several tens of nanometer of an epitaxial capping

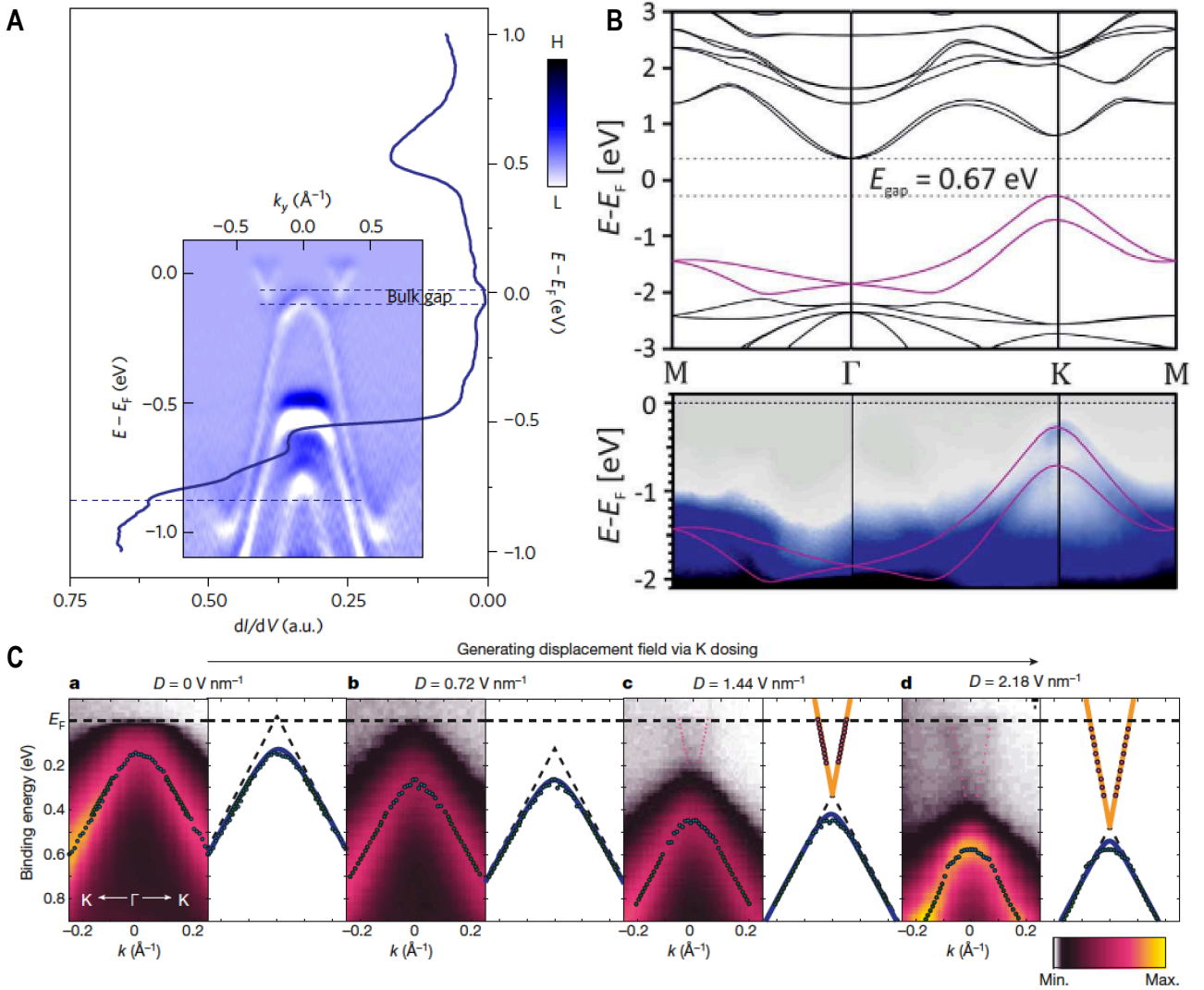

**Figure 7. Optical spectroscopy of QSH materials.** (A) (Inset) ARPES band structure measurements of 1T'-WTe$_2$ along the Γ-Y crystal direction (from [161]) compared to scanning probe spectroscopy to determine the bulk energy gap for the first time in an atomically thin quantum spin Hall insulator. Reproduced with permission.[161] Copyright 2017, Springer Nature. (B) (Top) DFT band structure calculation of atomically-flat bismuthene on substrate predicting an indirect bandgap of ~670 meV and a Rashba-split valence band at K. (Bottom) ARPES band structure measurement of atomically flat bismuthene on silicon carbide (0001) resolving a clear band gap and a Rasha splitting at the K point. Reproduced with permission.[52] Copyright 2017, AAAS. (C) ARPES measurements of Na$_3$Bi, along Γ-K, showing Fermi level shifts and gap narrowing as a results of charge transfer and a displacement field generated by in-situ potassium (K) doping. Reproduced with permission.[53] Copyright 2018, Springer Nature.

layer,[17] hindering ARPES access. This challenge is naturally overcome in atomically thin QSH

insulators in which the surface is usually exposed, allowing access by surface sensitive

techniques such as ARPES.[52, 53, 107, 117-119, 152, 161]

Whilst several authors have used ARPES to resolve the band structure of QSH candidate Xenes,[119, 128, 136-140, 143-145, 195-198] including isolated reports of sizable bulk bandgaps,[117, 118] confirmation of an edge state via an alternate experimental is usually required in order to confirm a QSH phase. The first ARPES measurements of a QSH insulator was reported only in 2017 by Tang et al.[161] for monolayer 1T'-WTe$_2$, who were able to resolve both electron and hole pockets close to the Fermi edge, offset in momentum space along the Γ-Y direction [161]. From a direct comparison of ARPES data with scanning tunneling spectroscopy (see **Figure 7A**) a band gap of 55 meV could be confirmed. Similar ARPES measurements have since been reported also for the related 1T'-WSe$_2$ compound[166, 167] ($E_g$~100 meV).

In the Xene group of materials, Deng et al.[107] have later resolved the band structure of monolayer stanene on Cu(111) substrates and measured a topological gap of ~300 meV at the Γ-point, located ~1.25 eV below the Fermi energy. More recently, Reis et al.[52] were able to resolve the valence band edge of a Rashba-split energy band in the 2D bulk of ultra-flat, monolayer bismuthene on SiC(0001), located 200 meV below $E_F$. As reproduced in Figure 7B, the valence band edge forms the lower energy bound to a massive ~700 - 800 meV topological band gap, as verified using scanning probe spectroscopy and density functional theory calculations, which so far holds the record for largest fundamental gap in any QSH material.

An electrostatically tunable band gap with maximum exceeding 300 meV at the Fermi energy has been demonstrated by Collins et al.[53] who investigated the band structure of atomically thin

Na$_3$Bi. Using potassium (K) surface charge transfer doping, Collins et al. were able to tune the electrochemical potential in situ by as much as 400 meV (Figure 7C). A concomitant vertical displacement field created by the ionized K dopants was shown to further cause a band gap narrowing to below ~100 meV.[53] Together with low-temperature local probe spectroscopy (see Section 5.3 below), this measurement has formed the first experimental demonstration of an electric field tunable topological phase transition and thus presents a major advance towards the implementation of QSH materials as active channels in topological electronic devices (see Section 6.2.1).

Despite ARPES's great success in the study of topological phases of matter[34] – especially for atomically thin systems – it also has its fundamental limitations. ARPES is usually regarded a surface sensitive technique. The escape depth of photoelectrons is in the range of several nanometers (depending on excitation energy). However, the thickness of atomically thin materials is only in the range of ~1nm. The information extracted from ARPES therefore necessarily also contains photoemission intensity arising from the electronic structure of the substrate. This has become apparent e.g. in Ref. [107], where the ARPES signal from the QSH insulator stanene is nearly outweighed by contributions of the Cu(111) electronic structure, enhancing the degree of difficulty in interpreting ARPES data. Whilst this can be seen as a limitation, ARPES signals from the substrate may also be utilized to disentangle the influence of a stabilizing substrate on the electronic properties as done for both the Xenes stanene and bismuthene.[52, 107] Particularly in the case of bismuthene, ARPES was able to confirm band structure calculations for the combined effects of on-site and substrate induced Rashba spin-orbit coupling, both contributing to its immensely large band gap.

Possibly a more severe limitation of conventional ARPES is its inability to obtain band structure information of electronic states above the Fermi level. This particularly limits APRES measurements of band gaps positioned at or even above the Fermi energy. A possible way around his limitation can be two-photon photoemission spectroscopy (2PPE), in which short laser pulses populate unoccupied states before a second laser pulse excites electrons into the vacuum level, allowing unoccupied states to be visualized.[199]

Finally, the ARPES energy and angular resolution highly depends on crystal quality and size. The presence of atomic defects and surface contamination,[163] for example, as well as polycrystallinity[161, 193] may decrease the ARPES energy and angular resolution beyond what could otherwise be achieved by state-of the-art light-source and detector technology at cryogenic temperatures. Polycrystallinity with certain rotational symmetries of the individual crystal domains can still result in high resolution ARPES data,[161] but the polycrystallinity needs to be taken into account when interpreting the photoemission spectra to obtain the shape of the Fermi surface. The limited lateral real-space resolution also limits ARPES in attempts to resolve the band dispersion of topological edge states as these typically constitute only a marginal fraction of the total area sampled for the incident light. The photoemission intensity arising from an edge state can thus be expected to be very weak and outweighed by a strong signal from bulk or substrate bands. There has been considerable progress in the development of micro- and nano-ARPES for smaller spot sizes down to ~ 100 nm.[200] Takayama et al.[127] used spin-resolved ARPES to detect faint signatures of a spin-split band crossing in Bi(111) multi-bilayers, $E \approx 70$ meV below the Fermi energy and at $k \approx 0.7$ Å$^{-1}$ along the Γ-Y direction,[127] showing that spin-resolved APRES can in principle allow to observe edge state dispersions.

## 6.3. Scanning Tunneling Microscopy and Spectroscopy

Limitations of real-space resolution in ARPES may be overcome by complementing with the extremely high energy and spatial resolution offered by local probe spectroscopy at cryogenic temperatures. Scanning tunneling microscopy and spectroscopy (STM/STS) measures a current across a tunnel junction formed at an atomically-sharp metal probe, brought into close proximity (typically 1 nm or less) of a material's surface. If a bias voltage $V$ is applied across the junction, electrons can preferentially tunnel from the tip into empty states in the surface, or vice versa. The resulting current $I$ can be expressed as,

$$I = \frac{4\pi e}{\hbar} \int_{-\infty}^{\infty} |M|^2 \rho_S(E - eV) \rho_t(E) [f(E - eV) - f(E)] dE \qquad (4)$$

where $e$ is the electron charge, $\hbar$ is the reduced Planck constant, $M$ is the tunneling matrix element connecting electronic states in tip and sample, $V$ is conventionally the voltage bias applied to the sample, with the tip grounded, $\rho_s$ and $\rho_t$ are the respective the tip and sample density of states (DOS), and $f(E) = \left(1 + \exp\left[\frac{(E - E_F)}{k_B T}\right]\right)^{-1}$ is the Fermi-Dirac distribution.

At low measurement temperature ($k_B T \ll eV$), the junction differential conductance can be shown to be directly proportional to the sample's energy dependent local density of states (LDOS),

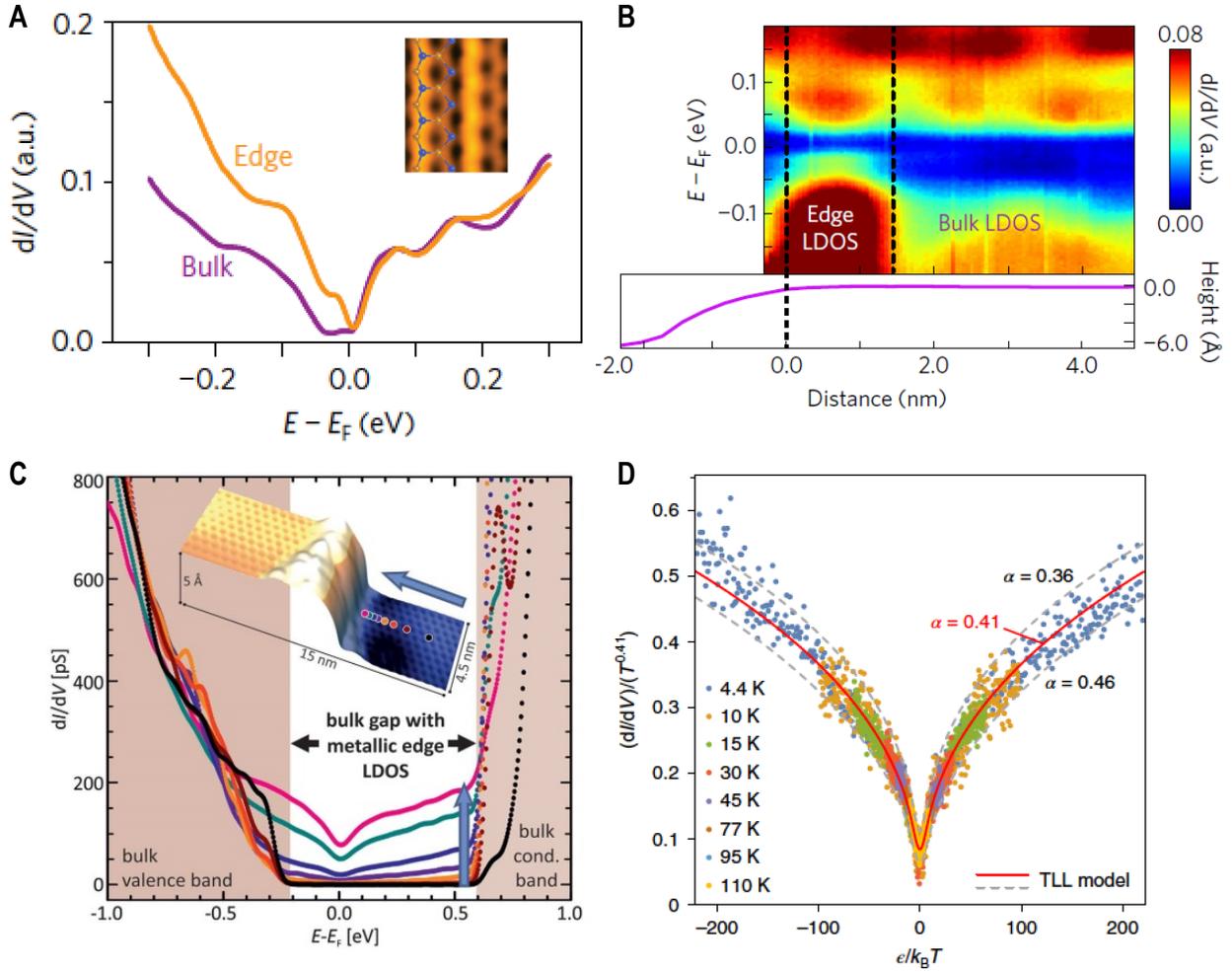

$$\frac{dI}{dV}(\vec{r}) \propto \rho_S(\vec{r}, E) = \sum_{n=0}^{\infty} \int |\psi_{nk}|^2 \delta(E - \epsilon_n(k)) dk \qquad (5)$$

**Figure 8. Scanning tunneling microscopy and spectroscopy measurements of QSH materials.** (A) Comparison of scanning tunneling spectra [161] of the local density of states (LDOS) acquired with the 2D bulk and at the edge of 1T'-WTe$_2$, showing a gapped bulk and conducting edges. (Inset) Topographic image of the related 1T'-WSe$_2$ [166]. Reproduced under the terms of the Creative Commons CC-BY-4.0 license.[166] Copyright 2018, The Authors, published by Springer Nature. (B) Spatial profile of the STS-resolved LDOS at the 1T'-WTe$_2$ edge [161], showing the transition from edge states to bulk states, compared a topographic height profile. (A, B) Reproduced with permission.[161] Copyright 2017, Springer Nature. (C) Similar data for bismuthene on SiC(0001).[52] (Inset) Topography of bismuthene on SiC(0001), showing the locations where the STS were acquired. Reproduced with permission.[52] Copyright 2017, AAAS. (D) Evidence of a helical Tomonaga-Luttinger liquid (TLL) formed at the QSH edge in bismuthene, demonstrating universal scaling of the edge-state LDOS [74] to extract the universal TLL power-law exponent $\alpha$. Reproduced with permission.[74] Copyright 2020, Springer Nature.

provided the tip DOS can be assumed featureless over the energy range of interest. Thus, STM/STS has direct access to the spatial dependence of electronic band gaps and metallic edge states[52, 161] via the quantum mechanical wave function detail, $|\psi_{nk}|^2$ at the atomic-level.

*5.3.1. Probing QSH gaps and edge states by STM/STS*

Although later interpreted to arise from the material's higher order topologically insulating (HOTI) nature[129], the first claim of 1D helical states resolved by STM/STS was obtained on Bi(111) bilayers by Kim et al.[142] and Drozdov et al.[126]. While STM had been used to study edges of Bi(111) as early 2012.[140], Kim et al., first observed the concurrence of a bandgap of $E_g$ ≈ 76 meV and a metallic edge state in the 2D bulk of a single Bi(111) bilayer on a $Bi_2Te_2Se$[142] consistent with bulk-boundary correspondence.[84, 86, 87] Drozdov et al., further demonstrated spectral differences between the two edge terminations on the surface layer of a (111) terminated bulk Bi sample, visualizing the HOTI's 1D helical hinge states.[126]

Representative bulk and edge spectra of a QSH helical edge were first demonstrated by Tang et al.[161] (**Figure 8A**). Sensitive measurements of the spatial dependence of the edge and bulk states are shown in Figure 8B,[161] where the transition between a gapped bulk and metallic edges that decay in to the 2D bulk can be observed. Directly comparing with ARPES (Figure 7A), allowed to determine a bulk bandgap of about 55 meV[161] in reasonable agreement with density functional theory predictions [44]. Similar measurements by Ugeda et al.[166] and Chen et al.[167] show slightly larger band gaps of $E_g$ ~ 0.1 eV on 1T'-$WSe_2$. Owing to well-defined crystalline phase boundaries between 2H- and 1T'-$WSe_2$ in the same crystals[166], a careful mapping allowed an extraction of the edge state decay length ~5 nm into the bulk. Similar edge and bulk properties have also been observed on small 1T'-MoS$_2$ islands,[165] though the extent to which quantum size effects influence the QSH gap in such small samples remains to be systematically investigated. As shown by Reis, et al., the transition from metallic edge states to a gapped bulk can be even more clearly observed in ultra-flat bismuthene (Figure 8C[52]), owing to bismuthene's enormous topological band gap ($E_g \approx 800$ meV) roughly centered at $E_F$ ($V = 0$ V).[52]

Perhaps one of the strongest claims of a QSH state was recently put forth by Collins et al. who have demonstrated control of the topological phase in monolayer Na$_3$Bi by vertical electric fields applied locally by an STM tip.[53] From STM/STS they observed QSH bandgaps as large as 360 meV in monolayer Na$_3$Bi that, when measured by STM/STS, proved susceptible to the STM probe's distance from the surface. Collins et al. argue that substantial built-in potential exists across the STM's tunnel junction arising from a difference in work function of tip metal and Na$_3$Bi. Such built-in potential results in an electric field that may be sensitively tuned by the tip-sample separation and may Stark shift bands and hence the topological band inversion. A completing closing of the topological gap at a certain critical field,[53] with subsequent reopening of the gap at higher electric field thus corroborates a topological phase transition.

*6.3.2. Probing superconductivity, quasiparticle interference and interactions by STM/STS*
Beyond precise measurements of bulk gaps in atomically thin QSH insulators, STM's extremely high spectral resolution also makes it particularly powerful in resolving the smallest energy gaps at cryogenic temperature, including superconducting quasiparticle gaps.[201-203] In the limit $k_B T \sim eV$, thermal broadening needs to be considered in the expression for the tunneling current and the differential conductance becomes,

$$\frac{dI}{dV}(\vec{r}) \propto \frac{df(E)}{dE}\rho_S(\vec{r},E) \qquad (6)$$

Limited by only thermal broadening, STS thus provides an energy resolution down to $3.5k_BT \approx$ 1.4 meV at $T \sim 4.5$ K, or even $3.5k_BT \approx 11$ μeV at $T = 38$ mK.[204] This particularly allows STM/STS to resolve QSH materials' non-trivial topology in the superconducting state [59] where it has very recently been able to confirm the emergence of Majorana bound states within the helical edge states of bismuth.[59]

Aside from real-space information, STM/STS also has unique access to electronic information in momentum space. Fourier Transform scanning tunneling spectroscopy (FT-STS) can resolve reciprocal space information by resolving local density modulations in real space, arising from quasiparticle scattering and interference. Analysis of such quasiparticle interference (QPI) patterns in Fourier space allows to reconstruct equi-energy surfaces at a given quasiparticle energy, similar to ARPES. Different from ARPES, however, QPI allows to obtain information of both filled and empty electronic states and may yield information of scattering probabilities and thus the preservation[205-207] of spin and pseudospin degrees of freedom.[147] Amongst 2D topological insulators, this technique was first applied to Bi(111) bilayers in which the band dispersion of the topological edge state could be confirmed [126], as well as to residual bulk states in $WTe_2$.[162] Song et al.[162] were able to resolve QPI in both electron and hole pockets at the Fermi level in 1T'-$WTe_2$ monolayers. In their samples, the two pockets appeared to be overlapping in energy with no apparent band gap. Recently, QPI was also used to observe the backscattering dispersion in a topological edge of finite length originating from ferromagnetic clusters localized to the ends of the edge,[147] thus demonstrating the spin-flip scattering enabled by TRS-breaking ferromagnetic clusters. Thus, QPI has proven a powerful tool in determining the band dispersion of 2D topological materials.

Finally, STM/STS has also been shown to detect the presence of strong Coulomb interactions in the QSH helical edge, in which the breakdown of Fermi liquid theory results in a highly correlated 1D electronic ground state – a helical Tomonaga-Luttinger liquid (TLL)[74] as detailed in Section 3. A TLL becomes apparent by a sharp suppression at zero bias ($E - E_F = 0$) (pseudogap) in the local density of state at the QSH helical edge and has frequently been observed in scanning probe based confirmations of the QSH state, e.g. in 1T'-TMDs,[161, 164-167] bismuthene,[52, 74] and $Na_3Bi$.[161] Most recently, a detailed study of the temperature and energy dependence of the zero-bias pseudo-gap feature on the edge of ultra-flat bismuthene was obtained by Stühler et al.[74] Figure 8D shows the temperature-dependence of the STS spectra, acquired in a temperature range from $T$ = 4.4 K to 110 K.[74] Traces of $dI/dV$ as functions of $eV$ and $T$ may be universally scaled and collapsed onto each other by dividing $dI/dV$ by $T^\alpha$ and $eV$ by $k_BT$ (Figure 3B,C), constituting the smoking gun signature of a TLL and allowing to extract the Luttinger parameter $K$.

*6.3.5. Probing Helicity by STM*

A remaining challenge in the study of 2D topological insulators is the unequivocal proof of spin-momentum locking, i.e. helicity at the edges of a QSH material. Theoretical proposals exist to detect helicity[208-211] by spin-polarized STM (SP-STM). SP-STM is based on tunneling of spin-polarized electrons from a magnetic probe tip, offering a high degree of spin polarization at

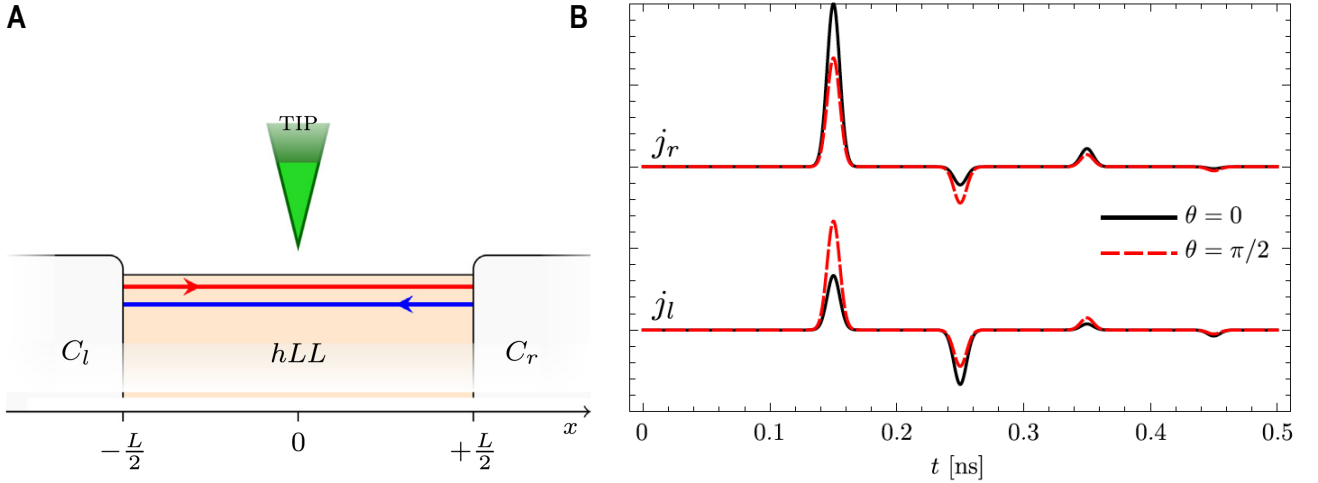

**Figure 9. Proposals to detect helicity in spin-polarized scanning tunneling microscopy.** (A) Schematic showing a majority spin population being injected into the helical Luttinger liquid (hLL) at the edge of a QSH insulator by a magnetic STM tip. The injected charge current fractionalizes and propagate via the 1D channels (red and blue) to be detected by the left and right contacts, $C_l$ and $C_r$. (B) Time-resolved spin currents $j_r$ and $j_l$ detected by the right and left contacts, respectively. Reproduced with permission.[210] Copyright 2016, Elsevier.

atomic resolution.[212] Thus, SP-STM may be used as a detection tool for a given spin-polarization or as a source to inject spin-polarized currents into a helical edge.[208, 210, 211]

Several publications have since proposed[208, 210, 211] that fractionalized charge and spin excitations may be observable in spatially-separated electrical contacts on either side of a spin-polarized probe tip localized along the QSH edge. A schematic[210] showing the proposed helicity detection method using injected spin-polarized electrons into a QSH edge is shown in **Figure 9A**. The difference in the tip and edge spin polarization, $\theta$, generates a fractionalized spin current that may be selectively detected by different electrical contacts along the edge, as spin-momentum locking preferentially directs the spin current towards one of the two terminals. Figure 9B illustrates the time resolution of this charge fractionalization effect for two different $\theta$, whereby time-resolved spin currents $j_r$ and $j_l$ are detected by the right and left electrodes, respectively.[210] Here, the largest peaks indicate the primary spin-polarized wave packet injected into the edge, and subsequent peaks and valleys indicate detection of spin packet reflections from the contacts.[210] To date, there have been no experimental confirmations of this approach in any QSH system, possibly owing to the materials and nanofabrication challenges to align electrical contacts to the QSH edge. Another reason may be that, in practice, the electrical contacts can drastically affect the observation of spin fractionalization.[210]

*6.3.5. Other Local Probe Techniques*

Aside from scanning tunneling microscopy, other local probe techniques have also been employed to investigate QSH systems, including scanning superconducting quantum interference device (SQUID),[69, 213] scanning gate microscopy,[70] and scanning microwave impedance microscopy.[39] Whilst these techniques usually lack the extremely high spatial resolution of STM, they can provide additional information such as spatially-resolved current flow to disambiguate whether current predominantly flows through the 2D bulk or through the edges.[69, 213] Complementary to STM, this allows to investigate electronic materials on a comparatively large scale up to ~100 μm, or when the QSH edge is protected by an insulating overlayer.[39]

**7. Potential Applications of Atomically thin QSH Materials**

The advent of 2D TIs offers device concepts that function radically differently from those based on semiconductors, presenting the opportunity to overcome some existing limitations of conventional semiconductor electronics. One key limitation in current semiconductor microelectronics is power dissipation, both in stand-by operation and during current switching, via subthreshold leakage currents through gate dielectrics or short channels and charging or discharging of capacitive loads, respectively.[214] Both dissipation sources could potentially be mitigated by alternative, non-charge-based switching mechanisms and by low-dissipation helical channels.

Modifying the electronic properties of a material for a targeted application is at the heart of microelectronics, which can be done in several ways. For example, in conventional semiconductors, one may use dopants to control charge carrier concentration and electronic band structure; external electric fields can dynamically control the band structure by band bending and charge accumulation or depletion; and even strain can be used to control charge carrier mobility. These control mechanisms are combined in metal-oxide-semiconductor field effect transistors (MOSFETs).[215, 216] where vertical electric fields induce charge at an oxide-semiconductor interface, gating current flow; and chemical doping defines metallic source and drain regions, allowing control over the electrostatic potential and carrier compensation in the channel region. Many of these concepts have been borrowed for QSH-based device proposals.[40, 43-46]

Beyond the bounds of classical electronics, the inclusion of materials with "non-classical" functionality, such as quantum mechanical phenomena like quantum coherence and entanglement, into electronic devices concepts can lead to entirely new paradigms superseding classical information processing.

## 7.1. Low-Power Electronics

Since the discovery of the quantum Hall effect in the 1980s[62] and its subsequent understanding as a topologically distinct phase of matter,[61] there has been excitement over the prospects for dissipationless (low-power) electronics. In QSH materials, such proposals[44, 45, 47] have their origin in the suppression of scattering predicted within the helical edge modes.

Classically, the amount of power dissipated in an electronic device, $P_D = VI = RI^2$, is related to the degree of charge carrier scattering, as the resistance $R$ is intimately tied to the carrier mobility in diffusive charge transport. However, within the QSH edge, carrier scattering should be strongly suppressed due to TRS-protected spin-momentum locking. Combined with a strictly 1D Fermi surface (consisting only of the two points at $+k_F$ and $-k_F$, respectively), momentum relaxation via single-particle backscattering requires a spin-flip that usually cannot occur in the absence of magnetic disorder or impurities.[26, 27] The extent to which scattering is suppressed within the QSH edge, how the edge state's resistance depends on its length,[17, 54, 55, 183] and which mechanisms contribute to backscattering are thus being heavily scrutinized.[17, 31, 54, 55, 186-189]

## 7.2. Topological "Switches": Control of the QSH State

There has been a lot of excitement over the prospects of manipulating the topological band structure of atomically thin QSH materials[17, 26, 27, 54, 55, 153] to realize topological switches that can be used as transistor-like devices[40, 43-47] for classical information processing.

State-of-the-art silicon MOSFETs are limited in their performance by a range of quantum and thermal effects, as well as capacitive and charge dynamics. One key performance metric of the MOSFET is their subthreshold swing, $S$, which is the sensitivity with which a transistor's drive current responds to an applied gate voltage at the turn-on. In a conventional transistor, this quantity is fundamentally limited by the thermal energy to $S = \ln(10)\frac{k_B T}{q} = 60$ mV/decade of current at room temperature.

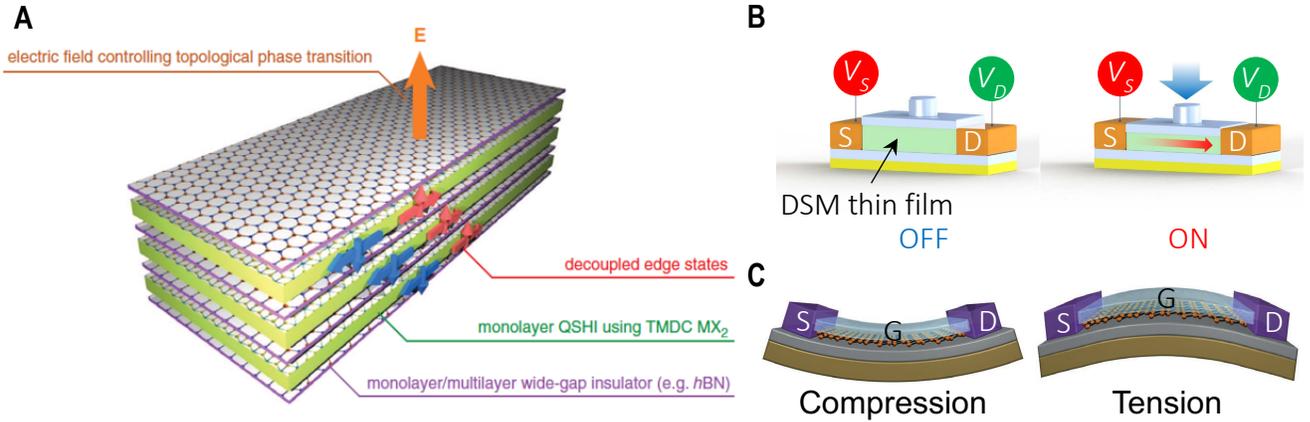

**Figure 10. External control of the QSH topological state.** (A) Proposal of a topological field-effect device [44], in which an externally applied vertical electric field E controls a topological phase transition from a QSH-insulator to trivial insulator, switching the 1D helical modes on and off. Reproduced with permission.[44] Copyright 2014, AAAS. (B-C) Schematics of strain-tunable QSH-insulators used as topological transistors. In (B), pressure applied to the basal plane of a thin Dirac semimetal (DSM). Adapted under terms of the Creative Commons CC-BY-4.0 license.[46] Copyright 2017, The Authors, published by Springer Nature. In (C), strain is imposed by flexure of a QSH material, which can apply both compressive and tensile strain. Adapted with permission.[43] Copyright 2017, Springer Nature.

Instead, reversible changes in the topology of atomically thin QSH materials may be achieved by strain and electric field, the latter of which may offer rapid controllability,[40, 44] allowing alternative ways to manipulate charge or spin currents.[45] In particular, a reversible phase transition from a topologically non-trivial to a trivial insulating or metallic state – a topological phase transition (TPT) – may be used to manipulate charge or spin currents[45] in the bulk or at the helical edge, and thus may find application as a current switch in a topological field-effect transistors,[40, 44, 47] with the potential to overcome fundamental limitations in conventional semiconductor electronics, including threshold slopes, switching speeds, and power dissipation.[53]

*7.2.1 Tunability by Electric Fields*

Different from the field effect in conventional semiconductor electronics, the "topological field effect"[40, 44, 47, 177] would rely on electric field control of the topological band structure by breaking inversion symmetry of the crystal lattice and/or change of the orbital energy by additional off- and on-site potentials. In certain atomically thin QSH systems, a vertical electric field can break inversion symmetry by changing the on-site potential when inverted orbitals are physically separated across different atomic planes.[45] This phenomenon usually precludes atomically flat or low-buckled materials such as the Xenes. If the applied electric fields are strong enough, a closing of the topological band gap and reopening of a topologically trivial gap may be achieved, constituting a topological phase transition, concomitant with the disappearance of the 1D topological edge state. As this "topological current switch" would be based on a phase transition rather than charge accumulation or inversion like MOSFETs, it may not to be fundamentally limited by thermionic emission across a potential barrier or to the same extent by capacitive delays.[45, 47]

**Figure 10A** shows and example of a topological field-effect device, based on stacked TMDC monolayers,[44] with a vertical gate electrode (related device proposals can be found in Ref's [40, 43, 47] for atomically thin TIs). In 1T'-TMDs, the transition metal *d* and the chalcogen *p* orbitals are separated across different vertical planes of the unit cell,[44] allowing field control of their electronic band structure as described above. For 1T'-$MoS_2$, for instance, Qian et al. predict critical electric fields of ~1.4 V $nm^{-1}$ to close the topological band gap and induce a topological phase transition, as shown in the phase diagram of **Figure 11A** and the band structure illustration in Figure 11B. Another example is the case of Na$_3$Bi monolayers, in which the Na and Bi $s$-orbitals and Bi $p_{x,y}$-orbitals are spatially separated,[53] and for which critical fields of $E_C = 18.5$ V nm$^{-1}$[53] have been predicted. The comparatively large predicted $E_C$ for Na$_3$Bi may simply reflect the high atomic defect density used in the model, which could screen an external electric field.[53]

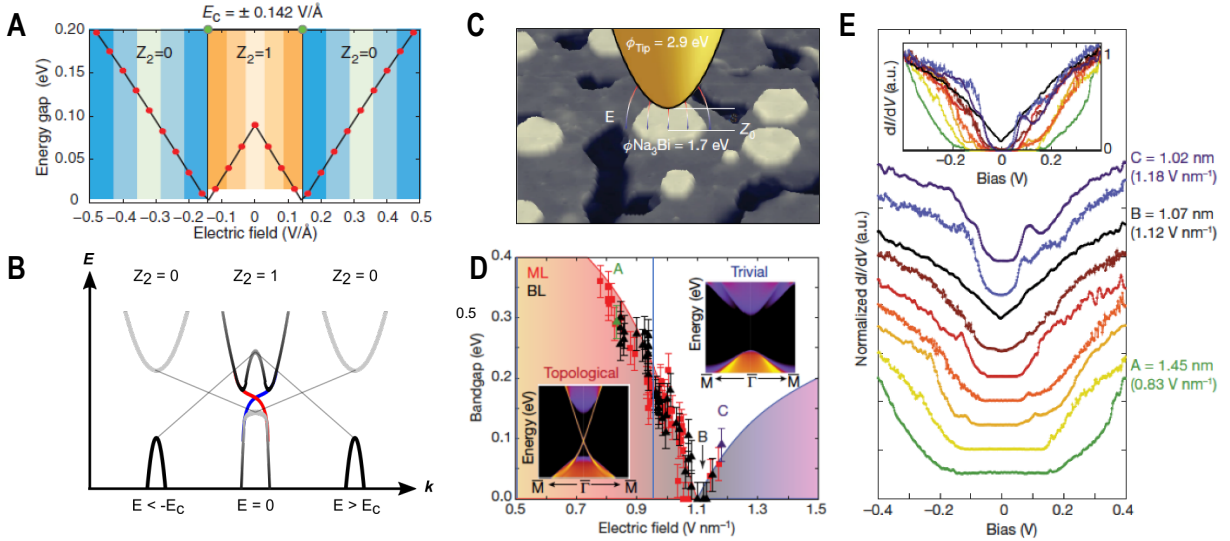

**Figure 11. Demonstration of the topological field effect.** (A) First-principles calculation of the 1T'-MoS$_2$ topological band gap as a function of a vertical electric field, predicting a topological phase transition from a topologically trivial ($Z_2 = 0$) to nontrivial ($Z_2 = 1$) insulator. Reproduced with permission,[44] Copyright 2014, AAAS. (B) Corresponding schematic band diagram of (A), showing band inversion and the emergence of helical edge states within the topological gap in the limit of no applied electric field ($E = 0$). (C-E) The topological field effect demonstrated in Na$_3$Bi. (C) A strong, built-in electric field is applied locally by an STM tip, given a mismatch of the sample and tip materials' work function difference $\Delta\phi$.[53] The electric field strength at the sample depends on physical separation $Z_0$ between tip and sample. (D) Measured topological phase diagram of monolayer and bilayer Na$_3$Bi as a function of electric field. (E) Measured local density of states within the 2D bulk from which the data in (D) was extracted, showing the closing of the topological gap (at low field strength) and the re-opening of a trivial gap (at high field strength). (C-E) Reproduced with permission.[53] Copyright 2018, Springer Nature.

Recently, Collins et al. were able to demonstrate a field-induced topological phase transition for the first time in any QSH material, applying estimated electric fields in the range of $E_C \approx 1.2 - 2.2$ V nm$^{-1}$, both globally via displacement fields from surface charge transfer doping, and locally with an STM probe.[53] The topological field-effect was detected by a change in the topological band gap, measured in ARPES, as well as by a complete closing and reopening of the gap, as seen in tip-height dependent local probe spectroscopy (see Figure 11C-E). A confirmation of such topological field effect in other QSH materials, as well as by electrostatic gate-control in an electronic device, will be major milestones towards QSH-based topological field-effect transistors.

*7.2.2. Tunability by Lattice Strain*

Tuning of the topological band structure by electric fields is limited to certain classes of QSH materials with vertically separated orbitals constituting the low-energy band structure. An alternative way to tune topological phase is by lattice strain. Atomic lattice constants govern the electronic band structure via the amount of spatial overlap between neighboring atomic orbitals, determining orbital energy, their hybridization, and, thus, the size of the fundamental band gap and degree of topological band inversion.[43, 89, 103, 217]

The effect of strain on the topological band structure was first demonstrated in HgTe/CdTe semiconductor heterostructure based QSH as an effort to enhance the size of the bulk band gap to increase the range of operating temperature.[68] For strained HgTe/CdTe heterostructures, the band gap was found to increase from $E_g \approx 10$ meV to 55 meV owing to compressive strain imparted by the growth substrate lattice.

Similar effects have since also been observed for atomically thin QSH systems, however, with much more profound effects. For instance, lattice strain imparted by the substrate can cause the Xenes to undergo a phase transition from a QSH to metal, such as buckled monolayer stanene on Sb(111)[105] or $Bi_2Te_3$[119] substrates. Recently, in monolayer 1T'-$WTe_2$,[169] the amount of substrate-induced strain was able to tune the material from a gapless topological semimetal to a fully gapped topological insulator with a bulk bandgap up to about 62 meV. Other QSH materials have also been predicted to be sensitive to strain, including all 1T'-TMDs,[44, 217, 218] Bi(110),[103] and low-buckled Xenes.[85, 90, 100, 121]

Strain-induced topological phase transitions between QSH and trivial insulator have also been predicted for larger-gap 2D TIs like functionalized, low-buckled stanene [90] and thin films of Weyl and Dirac semimetals, such as 1T'-TMDs,[44, 218] $Na_3Bi$,[46] and $Cd_3As_2$.[46] The susceptibility of the topological band structure to strain in these materials has led to proposals[43, 46] of a piezo-topological transistor, the schematics of which are shown in Figure 10B,C. In a piezo-topological transistor, switching is induced by dynamic strain applied to the channel of the transistor device by some external mechanism, which has included proposals of a mechanical pressure plate in physical contact with the QSHI channel (Figure 10B)[46] and by flexure of the entire device (Figure 10C).[43]

*7.2.3. Tunability by Magnetic Fields*

Magnetic fields universally break TRS and lead to the opening of a small Zeeman energy gap at the Kramers degenerate point (crossing point of the helical edge modes in the band structure) in crystals that lack additional protective symmetries.[16, 178] The concomitant change in the edge state conductance has been demonstrated in transport experiments,[17, 54, 55] where it has been provided as supporting evidence of the QSH state,[17, 54] and to determine the electron's spin-orbit enhanced *g*-factor.[54, 55]

If such TRS-breaking can be induced locally, it has the potential to create a localized "cut" the helical edge and define a set of two endpoints of the helical edge located on either side of the topologically trivial section. Combined with topological superconductivity, such a 1D system may carry Majorana bound states at the topological phase-boundary between trivial and helical parts of the edge,[219, 220] as discussed in the following section. With relevance to atomically thin QSH insulators in particular, magnetic fields may be locally imposed by ferromagnetic materials[59, 221] proximal to the QSH edge, such as ferromagnetic clusters[59] or even layered ferromagnets and ferromagnetic insulators,[221-228] the latter of which are fields that have burgeoned over the last three years. Layered ferromagnets exhibit intrinsic ferromagnetism in which each layer has a uniform magnetic polarization, even down to the monolayer limit, some of which are stable up to room temperature[223] and can be switched [223] between ferromagnetic and paramagnetic states with an applied gate voltage. A layered ferromagnet recently demonstrated[221] strong, polarization-dependent modulation of the edge conductance in the QSH material $WTe_2$, holding great promise as a "topological switch."

## 7.3. Spintronics

Spintronics aims to leverage the electron's spin degree of freedom instead of its charge for information processing and storage, and has been considered[229] for nonvolatility, fast processing speed, low power consumption, and large integration densities. A typical spintronic transistor or logic device is illustrated in **Figure 12A**, in which the spin channel contains a pure spin current, with opposing spin polarities propagating in opposite directions, as in the helical edge of a QSH insulator. Spin currents may generally be injected and detected through ferromagnetic or large atomic number (Z) metallic electrodes.

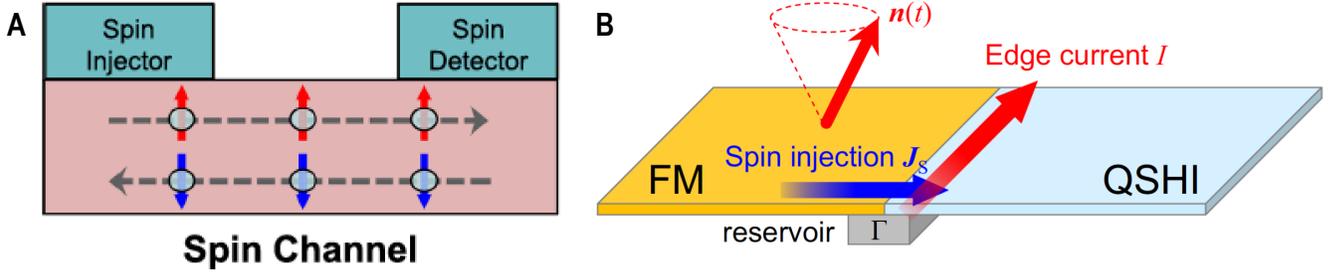

**Figure 12. Quantum spin Hall insulators for spin-charge conversion.** (A) Schematic of a QSH based spin transistor or logic device. Spin currents may generally be injected and detected through ferromagnetic or large atomic number (Z) metal electrodes. Ideally, electron spin is locked to its propagation direction within the spin channel. Reproduced under terms of the Creative Commons CC-BY-4.0 license.[41] Copyright 2018, The Authors, published by Springer Nature. (B) Schematic of spin current ($J_S$), generated from a QSH insulator edge that is exchange-coupled to an insulating ferromagnet (FM) having GHz time-dependent magnetization $n(t)$. The metallic reservoir (coupling parameter $\Gamma$) represents metallic terminals that would be coupled to the junction in a real setup, which allows the model system to maintain a periodic steady state. Reproduced under terms of the Creative Commons CC-BY-4.0 license.[42] Copyright 2020, The Authors, published by APS.

The efficient generation and detection of majority spin populations by interconversion of spin and charge is a major challenge in spintronics.[41] In 3D and 2D topological insulators, spin and charge are inherently linked via spin-momentum locking through their band dispersion properties, and may, thus, directly and efficiently convert between the two. The helical edges of QSH insulators in particular have been predicted to exhibit extremely efficient spin-charge interconversion.[42, 230] As the helical edge modes are spin-momentum locked, they may realize a pure spin current as illustrated in Figure 12B, where spin current ($J_S$) is injected into the QSH insulator via the magnetization dynamics of the ferromagnetic (FM) having a precessing (time-varying) magnetization $n(t)$, where it is converted to a charge current ($I$) on the edge of the QSH insulator edge. The metallic reservoir ($\Gamma$) represents metallic terminals that would be coupled to the junction in a real setup and allows the system to maintain a periodic steady state. Spins injected by a ferromagnet interfaced with a QSHI edge in this manner are expected[42] to be efficiently converted into charge, up to two orders of magnitude larger than in 3D topological insulator or Rashba interfaces, even despite suppression due to exchange coupling with the ferromagnetic terminals.[42] On the other hand, charge injected into the QSH insulator edge can be expected to perfectly convert to spin without Joule heating[42] due to the inherent spin-momentum locking within the edge channel.

Further detail on the field of spintronics and its applications, spin and charge interconversion by other quantum materials like 3D topological insulators, Rashba interfaces, ultra-thin Dirac and Weyl semimetals, superconductors, and non-collinear ferromagnets, as well as advances in 2D materials for spintronics applications can be found in these comprehensive reviews by Wolf et al.,[229] Han et al.,[41] Hirohata et al.,[230] Ningrum et al.,[231] and Ahn, E.[232]

### 7.4. Topological Superconductivity

Beyond the bounds of classical electronics, radically different applications of QSH materials may arise at the lowest extreme of temperatures, arising from the interplay of topology and superconductivity.

Soon after Kane and Mele's[15, 16] seminal proposal for the quantum spin Hall effect in graphene, Fu and Kane started exploring superconducting ground states in 2D and 3D topological insulators[56-58]. Superconducting pairing at the helical edge of a quantum spin Hall insulator, for example,[56, 219] was found to be a realization of Kitaev's model for 1D topological superconductivity,[60] with implications for the realization of proposals for Majorana based topological quantum information processing.[219, 233, 234]

Kitaev[60] first proposed harnessing Majorana fermions for solid state based quantum information processing in 2001. His model suggested the emergence of mid-gap (zero-mode) excitations that are localized at the ends of a finite 1D superconducting chain. Such *Majorana bound states* always occur in pairs – one on each end of the 1D chain – with their non-local 'entanglement' topologically protected from decoherence.[219, 220] This seemed to make Majorana fermions ideal candidates as carriers of quantum information and has made their realization ones of the most sought-after aspects in contemporary condensed matter research.[59, 235-249] However, in spite of its simplicity, a central assumption in Kitaev's model that made it hard to realize experimentally was the requirement for a spinless Hamiltonian. Furthermore, in 1D, superconducting correlations decay quickly as a power law, dissimilar to long-range correlations observed in three dimensional superconductors.

Just like ordinary Cooper pair wavefunctions in a superconductor, Majorana fermions can be described as superpositions of electron and hole states, or Bogoliubov quasiparticles. However, since the Bogoliubov quasiparticle operators are expressed as $\gamma = uc_\uparrow^\dagger + vc_\downarrow$ in spin singlet superconductors, the mixing of spin components in the fermion operators $c_\sigma$ implies that the Majorana condition $\gamma = \gamma^\dagger$ (a Majorana is its own antiparticle) is not satisfied. This limitation can be overcome in equal spin pairing, odd parity superconducting states where the Cooper pairs can be expressed in the so-called "spinless" form. However, materials showing an equal spin pairing, *p*-wave superconducting state are extremely rare.[250-252]

Alongside the emergence and increased theoretical understanding of topological matter around the turn of the century,[34, 35], it was shown that an effective *p*-wave Cooper pairing can be realized by proximity-coupling on the surfaces and interfaces of 2D and 3D topological insulators, where spin degeneracy is lifted by spin-momentum locking (helicity). In Fu and Kane's original proposal,[58] it was shown that Majorana fermions can be generated in cores of magnetic vortices by proximity coupling of a strong 3D topological insulator surface with an *s*-wave superconductor. Interestingly, such a junction could lead to formation of proximity induced, time reversal invariant, $p_x+ip_y$ superconductivity on the topological insulator even though the underlying superconductor has an *s*-wave superconducting state. Just a year later, Fu and Kane also considered 2D topological insulators[56, 57] such as quantum spin Hall insulators, in which the inherent lifting of spin degeneracy due to topological 'spin-filtering' within the helical edge naturally helps remove the fundamental 'fermion doubling'.[253]

Quite a few theoretical proposals have been put forth to discover Majorana fermions in quantum condensed matter (see Refs. [219, 220] for reviews). To date, Majorana fermions have been demonstrated in spin-orbit coupled semiconducting nanowires,[235-237, 249, 254, 255] atomic chains and island,[238, 239, 242, 256] the surface of iron-based superconductors,[245-248, 257] as well as 2D and 3D topological insulators.[59, 240, 244, 258] Amongst these systems, topological insulators may hold particular promise for applications as a topological non-trivial phase exists intrinsically, in the absence of an external magnetic field, with large bulk gaps, and robust to disorder.

Since the first demonstration of the QSH state in HgTe/CdTe and InAs/GaSb semiconductor heterostructures,[17, 23] a few reports on proximity-induced superconductivity in QSH systems have been published.[20, 259-262] The recent emergence of atomically thin QSH materials with intrinsic band inversion[44, 52-55, 161] and comparatively large topological gaps (a few tens to a few hundreds of meV) provides a fertile hunting ground for non-trivial superconductivity in these systems.

Particularly promising are the prospects of combining atomically thin QSH materials in van-der-Waals heterostructures with layered superconductors, ferromagnets or ferromagnetic insulators,[222-227] to explore the interplay of topology, superconductivity, and magnetism. Indeed, atomically thin layered materials have been shown to yield high-quality vdW heterostructures with atomically abrupt, crystalline interfaces,[168, 263-265] and with minimal strain.[266] These attributes make them promising platforms to achieve stable proximity-induced superconductivity.[168, 267-270] Recently, Lüpke et al. have demonstrated proximity-induced superconductivity in a superconducting vdW heterostructure of an atomically thin QSH material for the first time.[168]

Such ideas may by further combined with dynamical control of the topological state by local strain or electric fields[271-273] (see also references in Section 6.1 above). A very exciting prospect is the recent discovery of electric field controlled low-density superconductivity state in van-der-Waals heterostructures of $WTe_2$,[274, 275] which may ultimately allow for lateral homojunctions between QSH insulating and superconducting domains of the same material.

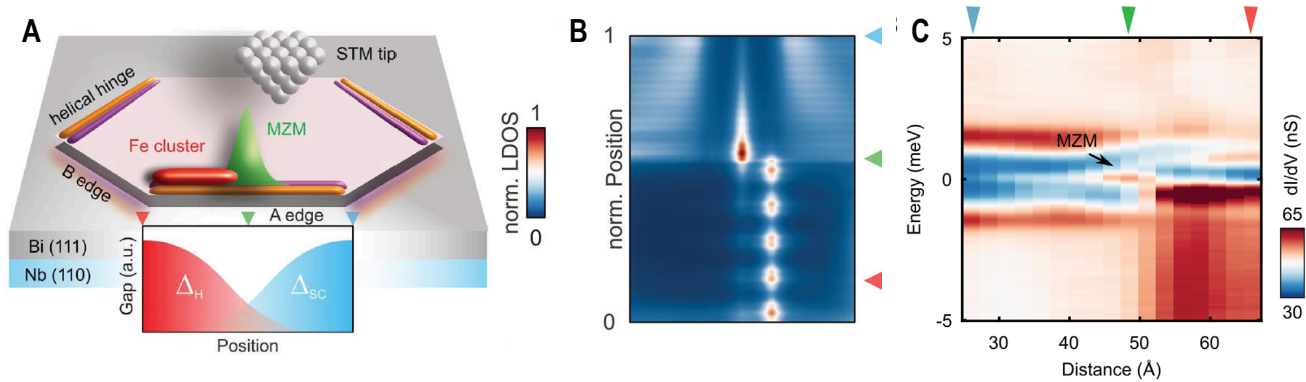

**Figure 13. Emergence and detection of Majorana zero modes at a superconducting helical edge.** (A) Similar to the helical edge of a QSH insulator, Majorana zero modes (MZM) can emerge at the superconducting helical "hinge" of a higher-order topological insulator in the vicinity of time-reversal symmetry breaking defects (e.g. clusters of magnetic atomic species). (B) Spatially-resolved, low energy LDOS calculation from a tight binding model across the interface illustrated in (A). (C) Spatially-resolved, STS-based detection of MZMs at the helical edge of the above heterostructure, evident by a zero-bias peak in the measured local density of states. The colored arrows in all panels correspond to the same features (red – hybridization gap, green – MZM, blue – superconducting gap). Reproduced with permission.[59] Copyright 2019, AAAS.

Majorana bound states are predicted to emerge at the boundaries between the 1D helical superconductor and a trivial insulator, defined either by physical end-points of a 1D wire or chain[235, 276] or by virtue of a locally controlled topological phase transition. Such bound states, or Majorana zero modes (MZM) may be detected in either electron transport[235] or local probe spectroscopy.[147, 238, 244] Although no report exists to date demonstrating Majorana bound states along the helical edge of a QSH insulator, a very recent report has demonstrated MZMs along the 1D helical hinge of the higher-order topological insulator (HOTI) Bi(111), proximitized by s-wave superconducting niobium.[59] A schematic of the heterostructure is shown in **Figure 13.** Here, the magnetic exchange fields originating from a small cluster of ferromagnetic atoms (Fe) along the helical edge breaks TRS locally, thus leading to the formation of strongly confined Majorana bound states that were predicted via tight binding calculation and detected in local probe spectroscopy at $T = 1.4$ K (see Figure 13). The local-probe detection of Majorana bound states in a 1D helical system promises similar discoveries in QSH materials in the future. Such discovery may ultimately provide a pathway to Majorana detection and braiding in QSH-based systems[233] towards scalable topological quantum computing.

## 8. Conclusion

A plethora of atomically thin QSH materials are now available to condensed matter research, promising transformative directions in materials and device engineering. This is owing to the stability and tunability of the topological state and the possibility to engineer novel electronic phases of matter, in a potentially scalable fashion, by van-der-Waals heterostructures with superconducting or magnetic layers, or both. As the most critical parameter, the spin-orbit driven QSH topological gap which limits a prospective device's operating temperature range keeps ever increasing as new materials are being discovered. With record values of topological band gaps in the range of up to few hundred meV – approaching the size of the silicon bandgap – this has brought even room temperature operation of QSH based electronics within reach.

Size matters. Yet, the range of potential applications of QSH materials are not restricted to the room temperature realm. The requirement of a large topological band gap is much relaxed for applications at cryogenic temperatures at which an even broader set of QSH materials with smaller topological gaps may find their place. In superconducting quantum devices, for instance, the operation temperature is necessarily limited to below the superconducting critical temperature. This has allowed topological energy gaps as small as ~1 meV[235] to be sufficient for the detection and control of topological superconductivity and Majorana fermions. The prospects of larger topological band gaps in atomically thin QSH materials that can both carry topological superconductivity in the absence of an applied magnetic field and be dynamically tuned by strain or electric fields is shaping up to be an exciting next frontier in QSH materials research.

**Acknowledgements**

This research is supported by the National Research Foundation (NRF) Singapore, under its Competitive Research Programme "Towards On-Chip Topological Quantum Devices" (NRF-CRP21-2018-0001). BW acknowledges a Singapore National Research Foundation (NRF) Fellowship (NRF-NRFF2017-11) and financial support from a Singapore Ministry of Education (MOE) Academic Research Fund Tier 3 grant (MOE2018-T3-1-002).

Received: ((will be filled in by the editorial staff))
Revised: ((will be filled in by the editorial staff))
Published online: ((will be filled in by the editorial staff))

**Conflict of Interest**

The authors declare no conflict of interest.